\documentclass[conference, compsocconf]{IEEEtran}
\usepackage{amssymb}
\usepackage{amsmath}
\usepackage{url}
\usepackage{graphicx}
\usepackage{subfigure}
\usepackage{epsfig}
\usepackage{times}
\usepackage{color}
\usepackage{multirow}
\usepackage{xspace}
\usepackage{afterpage}
\usepackage{epstopdf}
\usepackage{pifont}
\usepackage{pbox}
\graphicspath{{FIG/}}

\begin{document}

\newcommand{\hide}[1]{}
\newcommand{\parttitle}[1]{}
\newcommand{\note}[1]{{\textsf{\textcolor{red}{\ \ #1 \ }}}}
\newcommand{\jure}[1]{{{\textcolor{red}{[JL: #1]}}}}
\newcommand{\xhdr}[1]{\vspace{1mm}\noindent{{\bf #1.}}}
\newcommand{\eg}{\emph{e.g.}}
\newcommand{\ie}{\emph{i.e.}}
\newcommand{\model}{CESNA\xspace}
\newcommand{\fullmodel}{Communities from Edge Structure and Node Attributes\xspace}
\newcommand{\argmax}{\operatornamewithlimits{argmax}}
\newcommand{\argmin}{\operatornamewithlimits{argmin}}
\newcommand{\julian}{Circles\xspace}
\newcommand{\rev}[1]{\textbf{*** #1}}
\newcommand{\cmark}{\ding{51}}%
\newcommand{\xmark}{\ding{55}}%
\newcommand{\longver}[1]{#1}
\newcommand{\shortver}[1]{}



\title{Community Detection in Networks with\\Node Attributes}

\author{
\IEEEauthorblockN{Jaewon Yang}
\IEEEauthorblockA{
Stanford University\\
jayang@stanford.edu}
\and
\IEEEauthorblockN{Julian McAuley}
\IEEEauthorblockA{
Stanford University\\
jmcauley@cs.stanford.edu}
\and
\IEEEauthorblockN{Jure Leskovec}
\IEEEauthorblockA{
Stanford University\\
jure@cs.stanford.edu}
}

\maketitle

\begin{abstract}
%
Community detection algorithms are fundamental tools that allow us to uncover organizational principles in networks.
When detecting communities, there are two possible sources of information one can use: the network structure, and the features and attributes of nodes.
Even though communities form around nodes that have common edges \emph{and} common attributes, typically, algorithms have only focused on one of these two data modalities: \emph{community detection} algorithms traditionally focus only on the network structure, while \emph{clustering} algorithms mostly consider only node attributes.
In this paper, we develop \fullmodel (\model), an accurate and scalable algorithm for detecting overlapping communities in networks with node attributes. \model statistically models the interaction between the network structure and the node attributes, which leads to more accurate community detection as well as improved robustness in the presence of noise in the network structure. \model has a linear runtime in the network size and is able to process networks an order of magnitude larger than comparable approaches. Last, \model also helps with the interpretation of detected communities by finding relevant node attributes for each community.



\end{abstract}


\section{Introduction}
\label{sec:intro}

One of the most important tasks when studying networks is that of identifying \emph{network communities}. Fundamentally, communities allow us to discover groups of interacting objects (\ie, nodes) and the relations between them.
For example, in social networks, communities correspond to groups of friends who attended the same school, or who come from the same hometown~\cite{julian12circles}; in protein interaction networks, communities are functional modules of interacting proteins~\cite{Ahn10LinkCommunitiesNature}; in co-authorship networks, communities correspond to scientific disciplines~\cite{newman02community}.
Identifying network communities allows us to discover functionally related objects~\cite{Frank12MultiAssignmentClustering, newman02community, jaewon13agmfast}, study interactions between modules~\cite{airoldi07blockmodel}, infer missing attribute values~\cite{Cohen11BlockLDA, Coscia13DEMON}, and predict unobserved connections~\cite{RelationalLDA}.

\parttitle{How can we solve the problem? Node clustering, we can do on networks or attributes, or both!}

Identifying network communities can be viewed as a problem of clustering a set of nodes into communities, where a node can belong multiple communities at once.
Because nodes in communities share common properties or attributes, and because they have many relationships among themselves, there are two sources of data that can be used to perform the clustering task.
The first is the data about the objects (\ie, nodes) and their attributes. Known properties of proteins, users' social network profiles, or authors' publication histories may tell us which objects are similar, and to which communities or modules they may belong.
The second source of data comes from the network and the set of \emph{connections} between the objects. Users form friendships, proteins interact, and authors collaborate.


However, clustering methods typically focus only one of these two data modalities. In terms of attributes, \emph{clustering} algorithms~\cite{blei03lda,hierarchical} identify sets of objects whose attributes are similar, while ignoring relationships between objects. On the other hand, \emph{community detection} algorithms aim to find communities based on the network structure, \eg, to find groups of nodes that are densely connected~\cite{fortunato09community,Xie13SurveyOverlapping}, but they typically ignore node attributes.

By considering only one of these two sources of information independently, an algorithm may fail to account for important structure in the data. For example, attributes might tell us to which community a node with very few links belongs to; this would be difficult to determine from the network structure alone. Conversely, the network might tell us that two objects belong to the same community, even if one of them has no attribute information.
Thus, it is important to consider both sources of information together and consider network communities as sets of nodes that are densely connected, but which \emph{also} share some common attributes. Node attributes can complement the network structure, leading to more precise detection of communities; additionally, if one source of information is missing or noisy, the other can make up for it.
However, considering both node attributes and network topology for community detection is also challenging, as one has to combine two very different modalities of information.

Only recently have approaches for detecting communities based on both sources of information been developed~\cite{Cohen11BlockLDA,julian12circles} (Table~\ref{tab:baselines}). Many existing methods that combine network and node attribute information use single-assignment clustering~\cite{Akoglu12PICS, Ester06Cluster,Moser09Cluster,Ruan13CODICIL,Zhou09Clustering}; however, the applicability of these methods is limited, as they cannot detect overlapping communities.
Approaches based on topic models~\cite{Cohen11BlockLDA,LinkLDA, Sun12AttributeCluster,Xu12AttributedGraph} allow overlapping communities to be detected. However, they assume ``soft'' node-community memberships, which are not appropriate for modeling communities because they do not allow a node to have high membership strength to multiple communities simultaneously~\cite{jaewon11agmmodel2}.
Finally, all existing methods are only able to handle relatively small networks: the networks typically analyzed consist only of thousands of nodes~\cite{RelationalLDA,LinkLDA,julian12circles,Sun12AttributeCluster}.

\begin{table}[t]
\small
\begin{center}
\begin{tabular}{p{0.25\textwidth}cccc}
\hline
\pbox{0.25\textwidth}{Method class} & \pbox{0.07\textwidth}{$O$} & \pbox{0.07\textwidth}{$H$} & \pbox{0.09\textwidth}{$D$} & \pbox{0.09\textwidth}{$N$} \\ \hline
 {}Heuristics \cite{Akoglu12PICS,Ester06Cluster,Moser09Cluster,Ruan13CODICIL,Zhou09Clustering} & {{\xmark}} & {{\cmark}} & {{\xmark}} &  100,000\\ 
LDA-based \cite{Cohen11BlockLDA,RelationalLDA, LinkLDA, Sun12AttributeCluster, Xu12AttributedGraph} & {\cmark} & {\xmark} & {\cmark} & 85,000 \\ 
Clique-based heuristics\cite{Gunnermann13EDCAR,Gunnermann10Gamer} & {\cmark} & {\cmark} & {\xmark} &  100,000\\ 
Social circles \cite{julian12circles} & {\cmark}& {\cmark} & {\xmark} & 5,000 \\ 
 {\textbf{CESNA}} & \textbf{{\cmark}} & \textbf{{\cmark}} & \textbf{{\cmark}} & \textbf{1,000,000}\\
\hline
\end{tabular}
\end{center}
  \caption{Methods for community detection in networks with node attributes. $O$: Detects overlapping communities?, $H$: Assigns hard node-community memberships?, $D$: Allows for dependence between the network and the node attributes? (Fig.~\ref{fig:models}), $N$: Largest network that can be processed in 10 hours\longver{ (Fig.~\ref{fig:scalability}). Refer to Sec.~\ref{sec:related} for further details}.
  }
  \label{tab:baselines}
  \vspace{-7mm}
\end{table}

\begin{figure}[t]
\centering
  \subfigure[]{\label{fig:model.A}\includegraphics[height=0.1\textwidth]{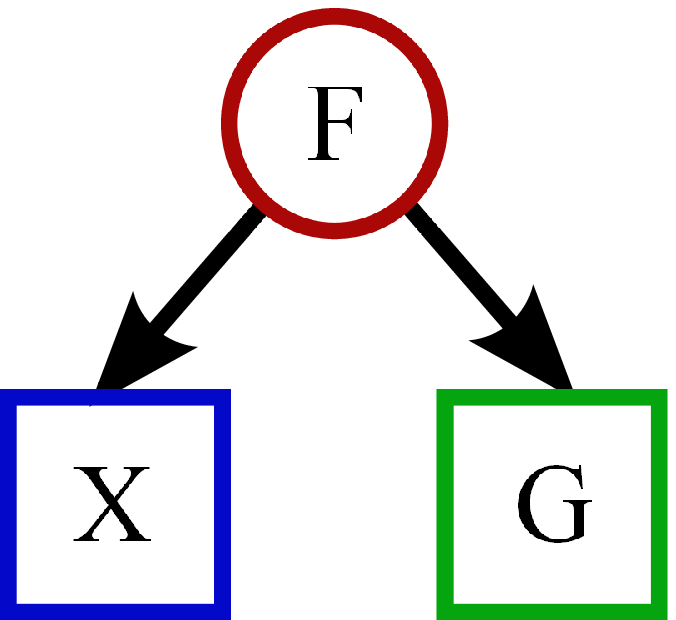}}
  \hspace{5mm}
  \subfigure[]{\label{fig:model.V}\includegraphics[height=0.1\textwidth]{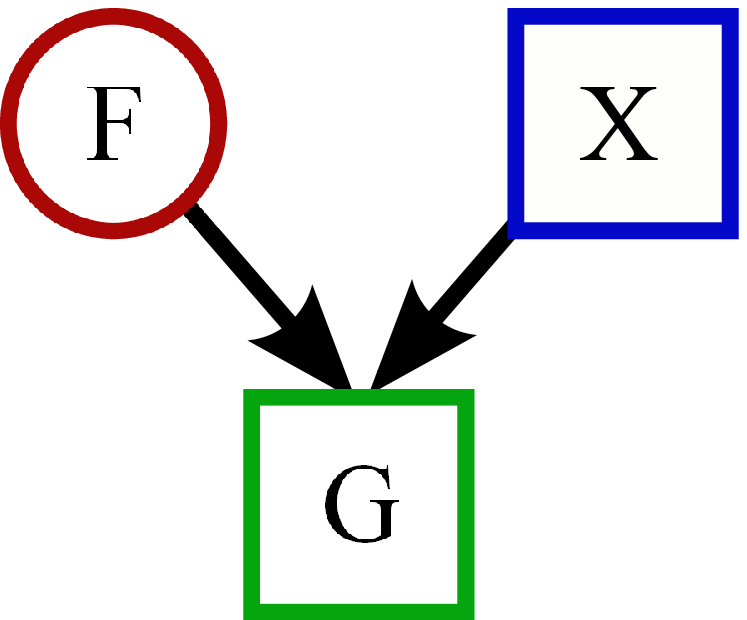}}
   \vspace{-2mm}
   \caption{
   Two ways of modeling the statistical relationship between a graph $G$, attributes $X$, and communities $F$. Circles represent latent variables that need to be inferred and squares represent manifest (observed) variables.
   }
   \label{fig:models}
   \vspace{-3mm}
\end{figure}

\xhdr{Present work: Community detection in networks with node attributes}
Here, we develop a high-performance (accurate and scalable) overlapping community detection method for networks with node attribute information. We present {\em \fullmodel} (\emph{\model}), which is based on a generative model for networks with node attributes. Our model advances existing approaches (summarized in Table~\ref{tab:baselines}) by making several innovations that ultimately lead to better performance both in terms of accuracy as well as scalability. First, our model allows us to detect overlapping communities by employing hard node-community memberships.
This way, we can avoid the assumption of soft-membership methods that nodes sharing multiple common communities are less likely to be connected~\cite{jaewon11agmmodel2}.
Second, in contrast to a line of previous work~\cite{Gunnermann13EDCAR,julian12circles}, which assumed that communities and attributes are marginally independent, we assume that communities ``generate'' both the network as well as attributes (Figure~\ref{fig:models}). This way we allow for dependence between the network and the attributes. Third, to fit the model and thus discover communities, we develop a block-coordinate ascent method where we can update all model parameters in time \emph{linear} in the number of edges in the network~\cite{jaewon13agmfast}. This makes our method scale to networks an order of magnitude larger than what was possible by previous methods.

To the best our knowledge, \model is the first overlapping community detection method that models both hard node-community memberships and the dependency between the communities and attributes. Moreover, \model can detect overlapping, non-overlapping, as well as hierarchically nested communities in networks, while considering \emph{both} node attributes and graph structure.

\hide{
We build a probabilistic generative model where we assign a non-negative membership factor between each node and each community.
Our model then assumes that these membership factors generate both the edges between the nodes and the node attributes.
To generate the network edges, we employ the generative process from BigCLAM and other affiliation network models~\cite{breiger74groups,feld86focused}, where the observed edges arise due to shared community affiliations between pairs of nodes.
To generate the node attributes, we consider a set of logistic models so that members of each community tend to share common traits.
} 

\hide{
\xhdr{Present work: Efficient and accurate community detection in networks with node attributes}
To identify communities from a given network with node attributes, we fit \model to the network and the attributes simultaneously, by identifying the most likely community memberships.
We develop a scalable, efficient fitting method based on block-coordinate ascent where we can update all model parameters in \emph{linear} time in the number of edges.
} 

We evaluate \model on six online social, information, and content-sharing networks: Facebook, Google+, Twitter, Wikipedia, and Flickr. We quantify \model's accuracy in detecting communities by comparing its predictions to hand-labeled ground-truth communities.
We compare \model to state-of-the-art community detection methods, including those that detect communities based only on the network structure, methods based only on node attributes, and methods that model both network structure and attributes jointly. Overall, \model achieves a 47\% improvement in the accuracy of detected communities over the baselines we consider.
We also examine whether node attributes can boost the performance of community detection algorithms in cases where the network is noisy or not fully observed. We add noise to the network and we find that the performance gap between \model over competing methods increases as the network structure becomes noisier and therefore less reliable.
This means that \model is able to successfully leverage node attributes to compensate for missing or noisy information in the network structure.

To quantify the scalability of \model we measure its running time on synthetic networks of increasing size. \shortver{(See the extended version~\cite{jaewon13attributes}.)} Compared to existing methods, the size of networks that \model can process far exceeds the current state-of-the-art: \model can handle networks 100 times larger than LDA-based methods~\cite{Cohen11BlockLDA} given the same runtime budget. Even when compared to methods that consider only the network structure (\ie, which handle strictly \emph{less} information), \model is faster than most baselines.

Last, we also inspect communities detected by \model on Facebook networks, and on a network of Wikipedia articles about famous philosophers. \shortver{(See the extended version~\cite{jaewon13attributes}.)}
We find that, on Facebook data, \model automatically learns that education-based attributes (``School name'' or ``Major'') are very highly correlated with a communities, whereas other people's attributes, such as ``Work start date'' and ``Work end date'' are not related to community structure.
%
On philosophers data, \model learns natural attributes for communities: \eg, subjects about Islamic culture are associated with a community of Islamic philosophers. While methods that ignore node attributes assign very influential philosophers (\eg, Aristotle) to most communities, \model circumvents this issue by modeling attributes, and discovering that Aristotle, while well connected to many philosophers, does not share common attributes with all of them.

\longver{
The rest of the paper is organized as follows. Section~\ref{sec:related} briefly surveys related work. In Section~\ref{sec:proposed}, we describe the statistical model of \model, and in Section~\ref{sec:fitting}, we discuss the parameter fitting procedure. We proceed by describing experimental evaluation in Section~\ref{sec:experiments} and conclude in Section~\ref{sec:conclusion}.
}

\hide{
One of the most important tasks in understanding networks is that of identifying sets of related nodes, or \emph{network communities}. Fundamentally, communities allow us to discover high-level structure among interacting objects. For example, in a social network, such high-level structures might be groups of friends who attended the same school, or who come from the same hometown~\cite{julian12circles}; in a protein interaction network, high-level structures might be functional modules of interacting proteins that share some common role~\cite{Ahn10LinkCommunitiesNature}; in a co-authorship network, high-level structures might be sets of authors who frequently collaborate~\cite{palla05_OveralpNature}. Identifying these structures allows us to identify users (or proteins etc.) that are similar to each other, to infer missing attributes, or to predict which connections are likely to form in the future.

There are two sources of data that are likely to help us to uncover such sets of related objects. The first source of data comes from the objects themselves, and their attributes. Known properties of proteins, users' social network profiles, or authors' publication histories might tell us which objects are similar, and to which communities or modules they may belong. The second source of data comes from the network, or the set of \emph{connections} between objects. Users form friendships, proteins interact, and authors collaborate.
In other words, nodes in communities are expected to share common properties or attributes, and also have many relationships among themselves.

Traditional methods for community detection focus on one of these two modalities. In terms of attributes, \emph{clustering} algorithms (such as K-means, Latent Dirichlet Allocation (LDA)~\cite{blei03lda}, or hierarchical clustering \cite{hierarchical}) identify sets of objects whose attributes are similar, while ignoring relationships between objects. On the other hand, \emph{community detection} algorithms aim to find communities based on the network structure, \eg, to find groups of nodes that are densely connected~\cite{fortunato09community,Xie13SurveyOverlapping}, but ignore node attributes.

However, by considering these two sources of information independently, we may fail to account for important structure in a network and its attributes. For example, attributes might tell us to which community a new user (\ie, a user who hasn't formed any friendships yet) belongs, which would be impossible from the network alone. Conversely the network might tell us that two users belong to the same community, even if one of them withholds their profile due to privacy reasons.

Ideally, then, we should consider both sources of information together. That is, we should consider network communities as sets of nodes that are densely connected, but which \emph{also} share some common attributes.
For example, proteins belonging to the same complex should possess the same chemical properties~\cite{Ahn10LinkCommunitiesNature}; individuals in a particular research community should share research interests~\cite{palla05_OveralpNature}; and members of a social group might have graduated from the same school, or come from the same hometown~\cite{julian12circles}.
Therefore, node attributes, if available, can complement the network structure, leading to more precise detection of communities; and if one source of information is missing or noisy, the other can make up for it.

\rev{
Only recently, a few approaches have been proposed to infer communities based on both sources of information~\cite{Cohen11BlockLDA,julian12circles}.
However, the existing methods have drawbacks in their modeling assumptions.
Many such methods are derived from topic models that assume ``soft-membership'', \ie, a node's membership is modeled in terms of its probability of belonging to a community
\cite{Cohen11BlockLDA,LinkLDA, Sun12AttributeCluster,Xu12AttributedGraph}.
In soft membership, high membership strength from a node to one community by definition precludes high membership strength from the node to other communities.
However, this assumption is no longer valid in other types of networks (\eg, in social networks), where nodes can have high membership strength to multiple communities simultaneously~\cite{julian12circles}.
}

\rev{
Instead of soft-membership, a few methods employ ``hard-membership'' where a node's community membership is independent each other~\cite{Ester06Cluster, julian12circles, Moser09Cluster, Ruan13CODICIL, Zhou09Clustering}. However, these methods have the other drawback that they do not model the direct relationship between communities and node attributes. In these methods, node attributes are assumed to be given \emph{independent} from community affiliations. In our experiments, however, modeling the relationship between communities and node attributes leads to better detection accuracy.
Finally, all existing methods are unable to handle large-scale networks: the networks analyzed typically consist of at most a few thousand nodes~\cite{julian12circles}.
}

\xhdr{Present work: A generative model to combine attributes and network structure}
\rev{
The goal of this paper is to develop a high performance (accurate and scalable) community detection method for networks where node attributes are also available.
To do this, we present \fullmodel (\model), a novel scalable method that is based on a generative model for networks with node attributes.
Our model advances the existing models by making the modeling choices that lead to better performance; Our model employs hard-membership rather than soft-membership, and it assumes that the node attributes are generated from communities rather than given independently. To the best our knowledge, \model is the first method that models both hard-membership and the communities-attributes relationship.
To maximize scalability, our model builds on BigCLAM~\cite{jaewon13agmfast}, which is the state-of-the-art community detection method both in terms of accuracy and scalability.
Our method possesses all the strengths of BigCLAM: it can detect overlapping, non-overlapping, as well as hierarchically nested communities in large-scale networks.
While BigCLAM is solely based on the graph structure, however, our method considers \emph{both} node attributes and graph structure, which leads to better performance.
}

We build a probabilistic generative model where we assign a non-negative membership factor between each node and each community.
Our model then assumes that these membership factors generate both the edges between the nodes \emph{and} the node attributes.
To generate the network edges, we employ the generative process from BigCLAM and other affiliation network models~\cite{breiger74groups,feld86focused}, where the observed edges arise due to shared community affiliations between pairs of nodes.
To generate the node attributes, we consider a set of logistic models so that members of each community tend to share common traits.

\hide{
Considering both node attributes and network topology for community detection is challenging as one has to combine two very different modalities of information. For example, iterative algorithms designed to operate on networks~\cite{palla05_OveralpNature,Ahn10LinkCommunitiesNature} do not provide an intuitive way to incorporate attributes.
A few approaches have been proposed that infer communities based on both sources of information~\cite{Cohen11BlockLDA,blei03lda,julian12circles}. Many such methods are based on topic models \cite{Cohen11BlockLDA,blei03lda}; however, these are designed to study ``topics'' among linked documents, which may be quite different from ``communities'' in other settings where we observe relational data. Furthermore, existing methods are unable to handle large-scale networks: the networks analyzed typically consist of at most a few thousand nodes~\cite{julian12circles}.

\xhdr{Present work: A generative model to combine attributes and network structure}
The goal of this paper is to detect communities in large-scale real-world networks where node attributes are also available.
To do this, we present \fullmodel (\model), a novel scalable method to detect communities in networks with node attributes.
We build on BigCLAM~\cite{jaewon13agmfast}, which is the state-of-the-art community detection method both in terms of accuracy and scalability.
Our method possesses all the strengths of BigCLAM: it can detect overlapping, non-overlapping, as well as hierarchically nested communities in large-scale networks.
While BigCLAM is solely based on the graph structure, however, our method considers \emph{both} node attributes and graph structure, which leads to better performance.

We build a probabilistic generative model where we assign a non-negative membership factor between each node and each community.
Our model then assumes that these membership factors generate both the edges between the nodes \emph{and} the node attributes.
To generate the network edges, we employ the generative process from BigCLAM and other affiliation network models~\cite{breiger74groups,feld86focused}, where the observed edges arise due to shared community affiliations between pairs of nodes.
To generate the node attributes, we consider a set of logistic models so that members of each community tend to share common traits.


Our generative model allows nodes to belong to multiple communities \emph{simultaneously} by modeling membership factors to each community as independent variables. This independent membership is a key difference from a large body of methods based on topic modeling~\cite{Cohen11BlockLDA,blei03lda} where a node's ``membership'' is modeled in terms of its probability of belonging to a community (a so-called ``mixed-membership''). Under such models, nodes cannot attain high membership weights to multiple communities simultaneously, as high membership strength to one community by definition precludes high membership strength to other communities. This makes sense when nodes represent ``documents'' and communities represent ``topics'', since (for example), a document that discusses both \emph{politics} and \emph{sport} presumably does not discuss sport with the same strength as a document that discusses sport alone. However this assumption is no longer valid in other types of network (\eg, in social networks), where nodes can have high membership strength to multiple communities simultaneously~\cite{julian12circles}.
}

\xhdr{Present work: Efficient and accurate community detection in networks with node attributes}
To identify communities from a given network with node attributes, we fit \model to the network and the attributes simultaneously, by identifying the most likely community memberships.
We develop a scalable, efficient fitting method based on block-coordinate ascent where we can update all model parameters in \emph{linear} time in the number of edges.

In our experiments on various social, information, and content-sharing networks with node attributes, we measure the accuracy of the detected communities compared to hand-labeled ground-truth. We compare \model to state-of-the-art community detection methods, including those that detect communities based only on the network structure, methods based only on node attributes, and methods that model both network structure and attributes jointly. Overall, \model achieves a 47\% improvement over the baselines we consider.
%
In another experiment where we add noise to the network, we find that the performance gap between \model over methods that rely on network information alone becomes larger as the network structure becomes noisier and therefore less reliable.
This result suggests that \model can successfully leverage node attributes to compensate for missing information from the network.

As a final quantitative evaluation, we measure the scalability of \model compared to various baselines by measuring their running time on synthetic networks of increasing size.
Compared to existing methods that model networks with node attributes, the size of networks that \model can handle far exceeds the current state-of-the-art:
\model can handle networks 100 times larger than topic models~\cite{Cohen11BlockLDA} given the same runtime budget.
Even compared to methods that consider only the network structure (\ie, which handle strictly \emph{less} information), \model is faster than most baselines, and is outperformed by the fastest (network-only) baseline by only 30\%.

\xhdr{Present work: Qualitative analysis}
To analyze the quality of the communities that \model detects in further detail, we inspect the detected communities on Facebook ego networks, and on a network of Wikipedia articles about famous philosophers (Philosophers).
For example, on Facebook data, \model automatically learns that education-based attributes (``School name'' or ``Major'') are very highly correlated with a few communities, whereas other features such as ``Work start date'' and ``Work end date'' are not related to community structure.
%
On Philosophers data, we examine the members of communities as well as their related attributes, and compare the results to methods that consider only the network structure.
Again, \model learns natural attributes for communities: \eg, subjects about Islamic culture are associated with a community of Islamic philosophers.
While methods that ignore node attributes assign very influential philosophers (\eg, Aristotle) to most communities, \model circumvents this issue by modeling attributes, and discovering that Aristotle, while well connected to many philosophers, does not share common attributes with all of them.
} 

\section{Related Work}
\label{sec:related}

We summarize the related work in Table~\ref{tab:baselines} and group it along two dimensions.
First, we consider how the methods model statistical dependency between communities, node attributes, and the underlying network (column $D$ of Table~\ref{tab:baselines}). Figure~\ref{fig:models} shows the two paradigms that are typically used. In Figure~\ref{fig:model.A}, community memberships $F$ generate both the graph $G$ and attributes $X$, while in Figure~\ref{fig:model.V}, $F$ and $X$ are given independently, and then the graph $G$ is generated by the interaction between $F$ and $X$.
Second, we focus on how the methods model the community memberships of individual nodes (columns $O$ and $H$).
Soft-membership models associate a probability distribution with the node's membership to communities, which means the more communities a node belongs to, the less it belongs to each individual community (simply because probabilities have to sum to one).
On the other hand, hard-membership models associate an independent binary variable for each node and community pair and, thus, do not suffer from the assumptions made by soft-membership models.


As shown in Table~\ref{tab:baselines}, heuristic single-assignment clustering methods for networks with node attributes~\cite{Ester06Cluster,Ruan13CODICIL,Zhou09Clustering} detect hard node-community memberships, however, because each node can belong to exactly one community, these methods cannot detect overlapping communities.

LDA-based methods~\cite{Cohen11BlockLDA,RelationalLDA,LinkLDA} aim to find sets of nodes that have similar ``topics'' of attributes and link among each other. These topic models are based on the paradigm in Figure~\ref{fig:model.A} where community memberships nodes generate links and node attributes. However, these methods assume soft community memberships, which leads to unrealistic assumptions about the structure of community overlaps~\cite{jaewon11agmmodel2}. We note that recently developed methods~\cite{Sun12AttributeCluster, Xu12AttributedGraph} also assume soft-membership and the paradigm in Fig.~\ref{fig:model.A}.

\hide{
However, soft-membership is not the best way to model community memberships.
In soft membership, high membership strength from a node to one community by definition precludes high membership strength from the node to other communities.
This makes sense when nodes represent ``documents'' and communities represent ``topics'', since (for example), a document that discusses both \emph{politics} and \emph{sport} presumably does not discuss sport with the same strength as a document that discusses sport alone.
However, this assumption is no longer valid in other types of networks (\eg, in social networks), where nodes can have high membership strength to multiple communities simultaneously~\cite{julian12circles}.
Indeed, these topic models have been outperformed by methods that detect hard memberships~\cite{julian12circles,jaewon13agmfast}. Therefore, \model aims to detect hard memberships also.

Recently, \cite{julian12circles} developed a pseudo-boolean method that detects overlapping communities with hard-membership.
This model is based on the paradigm in Figure~\ref{fig:model.V}.
However, we note that the paradigm in Figure~\ref{fig:model.A} is a better paradigm than the paradigm in Figure~\ref{fig:model.V} due to the following key observation.
Conceptually, Fig.~\ref{fig:model.A} assumes a direct connection from $F$ to $X$; however, Figure~\ref{fig:model.V} assumes $F$ and $X$ are {\em marginally independent}.
This difference has a large practical advantage for Fig.~\ref{fig:model.A};
if $G$ is noisy (for example) Fig.~\ref{fig:model.A} still can infer $F$ from $X$. On the other hand, Fig.~\ref{fig:model.V} does not allow $F$ to be inferred from $X$.
Thus, we choose the paradigm in Fig.~\ref{fig:model.A} for our model.

Table~\ref{tab:baselines} summarizes our discussion for related work.
In summary, our model differs from all the existing method as we adopt hard-membership, overlapping communities, and the paradigm in Figure~\ref{fig:model.A}.
Our experiments (Sec.~\ref{sec:experiments}) show that our design choices lead to improved performance.
}


\hide{
Identifying sets of related nodes in networks with node attributes has been studied by a few different lines of research.
Among the two sources of information (attributes and network structure), most traditional methods focus only on one of the two while ignoring the other.
For example, clustering methods~\cite{Frank12MultiAssignmentClustering,hierarchical} group nodes based only on their attributes.
On the other extreme is community detection~\cite{fortunato09community,Xie13SurveyOverlapping}, which is perhaps more closely related to our work. Community detection methods aim to find sets of densely connected nodes (\ie, communities) primarily based on the network structure.
%
%
There are a few community detection methods~\cite{julian12circles,Ruan13CODICIL} that can handle networks with node attributes like our work here.
However, these methods do not capture the connection between the attributes and the communities because they assume that the node attributes are given independently from communities. On the other hand, our model assumes that the node attributes are the results of community memberships of the nodes.




Another kind of approaches that exploit both attributes and the network structure is based on topic models~\cite{Cohen11BlockLDA,RelationalLDA,LinkLDA}, which aim to find sets of documents that discuss the same topic based on word occurrences and links between documents.
By representing nodes as documents and attributes as words, these models can be directly applied to the problems we investigate here.
However, there is a crucial difference between modeling communities versus topic memberships of nodes.
These models associate a probability with each node and community (topic), \ie, each node has a ``soft membership'' to all communities.
In practice, models that generate soft memberships~\cite{airoldi07blockmodel,Barbieri13CascadeCommunity, Sun12AttributeCluster} differ from models that allow nodes to belong to multiple communities \emph{simultaneously}. Topic modeling approaches are generally designed to model topic assignments of \emph{documents}, which may be quite different from community memberships of users in a social network (for example). Indeed, these topic models have
been outperformed by methods that detect ``hard'' memberships~\cite{julian12circles,Ruan13CODICIL}. Therefore, \model aims to detect hard memberships also.


Finally, we briefly note that studies on modeling networks such as Infinite Relational Models~\cite{Miller09InfiniteRelationalModel,morup11infinite} or Multiplicative Attribute Graphs~\cite{Myunghwan12MultiGroup} can be applied to networks with node attributes. However, these models have a different goal than our work, as they aim to model and predict link formation between nodes.
}

\section{\model Model Description}
\label{sec:proposed}

Here, we develop a probabilistic model that combines community memberships, the network topology, and node attributes. We present the {\em \fullmodel} (\model), a probabilistic generative model for networks and node attributes that satisfies the desiderata mentioned above. Our model is based on the following intuitive properties:
\begin{itemize}
\item Nodes that belong to the same communities are likely to be connected to each other.
\item Communities can overlap, as individual nodes may belong to multiple communities.
\item If two nodes belong to multiple common communities, they are more likely to be connected than if they share only a single common community (\ie, overlapping communities are denser~\cite{feld86focused,jaewon11agmmodel2}).
\item Nodes in the same community are likely to share common attributes --- for example, a community might consist of friends attending a same school.
\end{itemize}

We formally describe the generative process of \model as follows. We assume that there are $N$ nodes  in the network $G$, each node has $K$ attributes, and there are $C$ communities in total.
We denote the network by $G$, the node attributes by $X$ ($X_{uk}$ is $k$-th attribute of node $u$), and community memberships by $F$. For community memberships $F$, we assume that each node $u$ has a non-negative affiliation weight $F_{uc} \in [0, \infty)$ to community $c$. ($F_{uc} = 0$ means that node $u$ does not belong to community $c$.)

We shall now proceed by describing these components of the model in further detail.

\xhdr{Modeling the links of the network}
To model how the network structure depends on node community memberships, we aim to capture the following three intuitions:
\begin{enumerate}
\item node community affiliations influence the likelihood that a pair of nodes is connected,
\item the degree of influence (the probability that nodes belonging to the same community are connected) differs per community, and
\item each community influences this connection probability independently.
\end{enumerate}

To achieve these goals, we build on Affiliation Network Models~\cite{breiger74groups, feld86focused, Lattanzi09AffiliationNetworks, jaewon11agmmodel2}, where the graph $G(V, E)$ arises from node community memberships $F$. To generate the adjacency matrix $A \in \{0, 1\}^{N \times N}$ of network $G$, we employ the probabilistic generative process of the BigCLAM overlapping community detection algorithm~\cite{jaewon13agmfast}. In particular, we assume that two member nodes $u, v$ belonging to a community $c$ are connected with the following probability:
\[
P_{uv}(c) = 1 - \exp(-F_{uc} \cdot F_{vc}).
\]
Note that if either $u$ or $v$ does not belong to $c$ ($F_{uc} = 0$ or $F_{vc} = 0$), these nodes would not be connected ($P_{uv}(c) = 0$).

We assume that each community $c$ connects nodes $u, v$ independently with probability $1 - \exp(-F_{uc} \cdot F_{vc})$. From this, we can derive the edge probability $P_{uv}$ between nodes $u$ and $v$. In order for $u, v$ to be unconnected, the nodes $u$ and $v$ should not be connected in \emph{any} community $c$:
\[
1 - P_{uv} = \prod_c (1 - P_{uv}(c)) = \exp ( - \sum_c F_{uc} \cdot F_{vc}).
\]
In summary, we assume the following generative process for each entry $A_{uv} \in \{0, 1\}$ of the network's adjacency matrix:
\begin{equation}
\begin{split}
P_{uv} & =   1 - \exp(- \sum_c F_{uc}\cdot  F_{vc}),\\
A_{uv} & \sim  \mbox{Bernoulli}(P_{uv}).
\end{split}
\label{eq:puv}
\end{equation}

Note that the above generative process satisfies our three aforementioned requirements. The network edges are created due to shared community memberships (Requirement (1)). Furthermore, each membership $F_{uc}$ of a node $u$ is regarded as an independent variable to allow a node to belong to multiple communities simultaneously (Requirement (2)). This is in stark contrast to ``soft-membership'' models (such as mixed membership stochastic block models~\cite{airoldi07blockmodel}), which add constraints $\sum_c F_{uc} = 1$ so that $F_{uc}$ is a probability that a node $u$ belongs to a particular community. Finally, because each community $c$ generates connections between its members independently, nodes belonging to multiple common communities have a higher probability of connecting than if they share just a single community (Requirement (3)).

\xhdr{Modeling node attributes}
Just as community affiliations can be used to model network edges, they can also be used to model node attributes. We next describe how node attributes are generated from community memberships.

We assume binary-valued attributes where for each attribute $X_{uk}$ of a node $u$, we consider a separate logistic model. Our intuition is that, based on a node's community memberships, we should be able to predict the value of each of the node's attribute values. Thus, we regard group memberships $F_{u1}, \ldots, F_{uC}$ as input features of the logistic model with the associated logistic weight factor $W_{kc}$ (for each attribute $k$ and community $c$). We also add an intercept term $F_{u(C + 1)} = 1$ to the input feature of each node $u$:
\begin{equation}
\begin{split}
Q_{uk} &= \frac{1}{1 + \exp( - \sum_{c} W_{kc} \cdot F_{uc})},\\
X_{uk} & \sim  \mbox{Bernoulli}(Q_{uk})
\end{split}
\label{eq:xuk}
\end{equation}
where $W_{kc}$ is a real-valued logistic model parameter for community $c$ to the $k$-th node attribute and $W_{k(C+1)}$ is a bias term. The value of $W_{kc}$ represents the relevance of each group membership $c$ to the presence of a particular node attribute $k$.

\begin{figure}[t]
  \centering
  \includegraphics[width=0.37\textwidth]{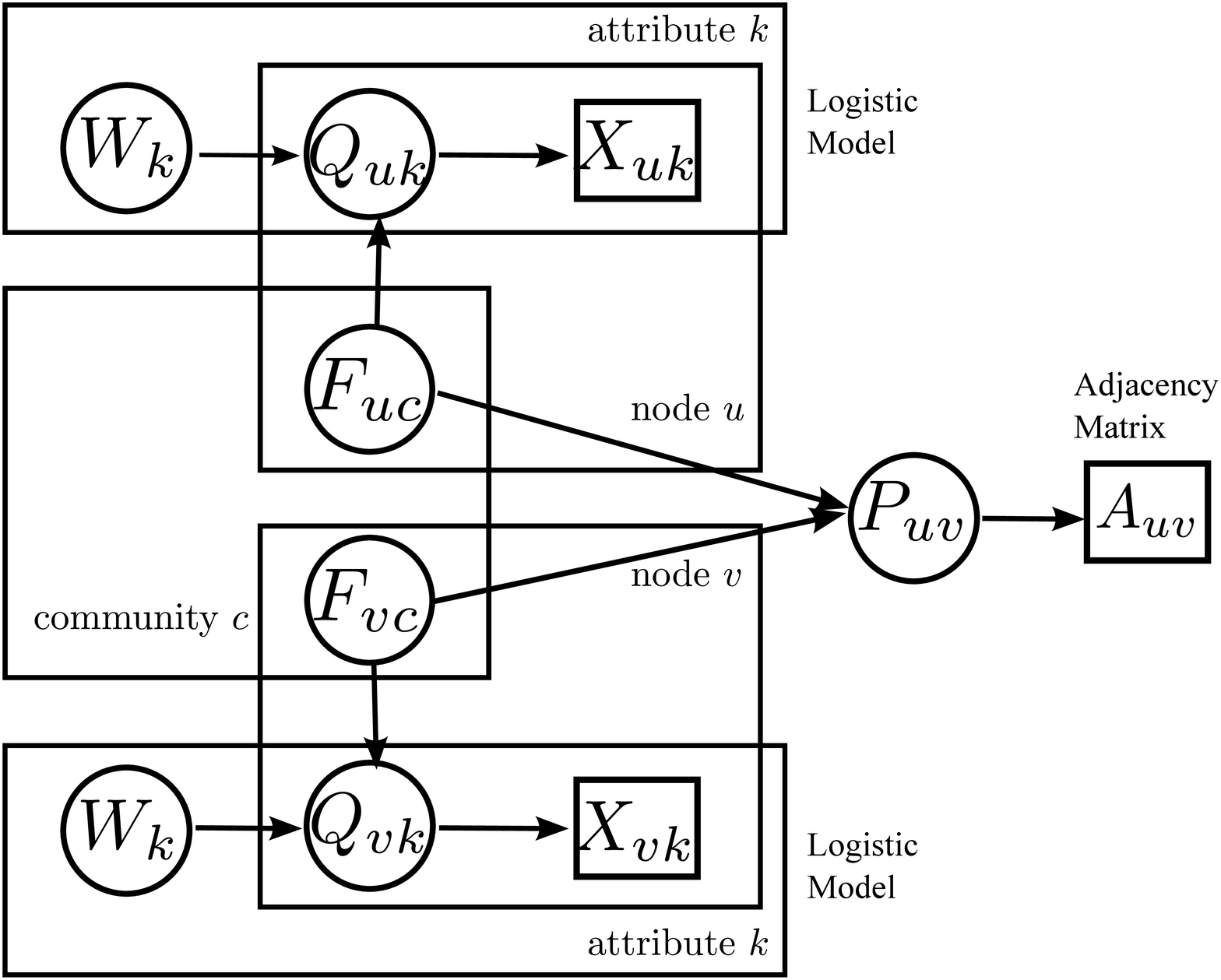}
 \vspace{-2mm}
   \caption{
   Plate representation of \model. $X_{uk}$: $k$-th attribute of node $u$; $W_k$: Logistic weight vector for attribute $k$; $Q_{uk}$: Probability that $X_{uk} = 1$; $F_{uc}$: Membership strength of node $u$ to community $c$; $A_{uv}$: Indicator for whether the nodes $u$ and $v$ are connected; $P_{uv}$: Probability that $A_{uv} = 1$.
  }
  \label{fig:plate}
\vspace{-6mm}
\end{figure}

Figure~\ref{fig:plate} illustrates the \model model. Rectangles ($X_{uk}$, $A_{uv}$) are the node attributes and the network adjacency matrix that we observe. Circles denote latent variables: community memberships $F$ and logistic weights $W$. We explain how to estimate community memberships from node attributes and the network structure (\ie, how we infer $F$ from $X$ and $A$) in the following section.

Last, we also note that depending on the type of attribute, there are also other choices for modeling attributes $X$ based on $F$. For example, for real-valued attributes linear regression could be used.
Also, note that we assume that the number of attributes is relatively small compared to the number of nodes; as such, we can use a separate logistic model for each attribute. In the case of many attributes, one could consider methods that group attributes as well as nodes~\cite{Myunghwan12MultiGroup}.

\section{Inferring Communities with \model}
\label{sec:fitting}

Next, we shall describe how we detect network communities by estimating \model model parameters from given data. We are given an undirected graph $G(V,E)$ with binary node attributes $X$. We aim to detect $C$ communities as well as the relation between communities and attributes. For now, we shall assume the number of communities $C$ is given. Later, we will describe how to automatically estimate $C$.

We aim to infer the values of latent variables $F$ and $W$ based on the observed network and the attributes. This means we need to estimate $N \cdot C$ community memberships (\ie, $\hat{F} \in \mathbb{R}^{N \times C}$), and $K \cdot (C + 1)$ logistic weight parameters (\ie, $\hat{W} \in \mathbb{R}^{K \times (C + 1)}$).

We find the optimal $\hat{F}$ and $\hat{W}$ by maximizing the likelihood $l(F, W) = \log P(G, X|F, W)$ of the observed data $G, X$:
\begin{equation}
\hat{F}, \hat{W} = \argmax_{F \geq 0, W} \log P(G, X|F, W).
\label{eq:mle}
\end{equation}

Because $G$ and $X$ are conditionally independent given $F$ and $W$, we can decompose the log-likelihood $\log P(G, X|F, W)$ as follows:
\[
\log P(G, X|F, W) = \mathcal{L}_G + \mathcal{L}_X
\]
where $\mathcal{L}_G = \log P(G|F)$ and $\mathcal{L}_X = \log P(X|F, W)$. We compute $\mathcal{L}_G$ and $\mathcal{L}_X$ simply using Equations~\ref{eq:puv} and ~\ref{eq:xuk}:

\[
\begin{split}
\mathcal{L}_G & = \sum_{(u, v) \in E} \log (1 - \exp( - F_{u} F_{v}^T)) - \sum_{(u, v) \not\in E} F_{u} F_{v}^T\\
\mathcal{L}_X & = \sum_{u, k} (X_{uk} \log Q_{uk} + ( 1 - X_{uk}) \log (1 - Q_{uk})),
\end{split}
\]
where $F_u$ is a vector $\{F_{uc}\}$ for node $u$ and $Q_{uk}$ is defined in Equation~\ref{eq:xuk}.

Last, we also invoke $l_1$-regularization on $W$ to avoid overfitting and to learn sparse relationships between communities and attributes. Thus, our optimization problem that we aim to solve is:
\begin{equation}
\hat{F}, \hat{W} = \argmax_{F \geq 0, W} \mathcal{L}_G + \mathcal{L}_X - \lambda |W|_1,
\label{eq:mle_full}
\end{equation}
where $\lambda$ is a regularization hyperparameter.

To solve the problem in Eq.~\ref{eq:mle_full}, we adopt a block coordinate ascent approach. We update $F_u$ for each node $u$ by fixing both $W$ and the community membership $F_v$ of all other nodes $v$. After updating $F_u$ for all nodes, we then update $W$ while fixing community memberships $F$. This way, we can decompose the non-convex optimization problem of Eq.~\ref{eq:mle_full} into a set of \emph{convex} subproblems. We describe our solution to each of these subproblems next.

\xhdr{Updating community memberships}
To update community memberships, we build on the optimization procedure used in BigCLAM~\cite{jaewon13agmfast}. 
However, we modify the procedure to consider node attributes (which BigCLAM ignores). We update the membership $F_u$ of an individual node $u$ while fixing all other parameters (the membership $F_v$ of all other nodes, and logistic model parameters $W$).

We solve the following subproblem for each $u$:
\begin{equation}
\hat{F_u} = \argmax_{F_{uc} \geq 0} \mathcal{L}_G(F_u) + \mathcal{L}_X(F_u),
\label{eq:mle_onenode}
\end{equation}
where $\mathcal{L}_G(F_u)$ and  $\mathcal{L}_X(F_u)$ are the parts of $\mathcal{L}_G, \mathcal{L}_X$ involving $F_u$, \ie,
\[
\begin{split}
\mathcal{L}_G(F_u) &= \sum_{v \in \mathcal{N}(u)} \log (1 - \exp( - F_{u} F_{v}^T)) - \sum_{v \not\in \mathcal{N}(u)} F_{u} F_{v}^T \\
\mathcal{L}_X(F_u) &= \sum_{k} (X_{uk} \log Q_{uk} + ( 1 - X_{uk}) \log (1 - Q_{uk}))
\end{split}
\]
where $\mathcal{N}(u)$ is a set of neighbors of $u$. Note that this problem is convex: $\mathcal{L}_G(F_u)$ is a concave function of $F_u$~\cite{jaewon13agmfast,morup11infinite} and $\mathcal{L}_X(F_u)$ is a logistic function of $F_{uc}$ when $W$ is fixed.

To solve this convex problem, we use projected gradient ascent. The gradient can be computed straightforwardly:
\[
\begin{split}
\frac{ \partial \mathcal{L}_G(F_{u})} {\partial F_u} & = \sum_{v \in \mathcal{N}(u)} F_{vc} \frac { \exp( - F_{u} F_{v}^T)} {1 - \exp( - F_{u} F_{v}^T)} - \sum_{v \not\in \mathcal{N}(u)} F_{vc}\\
\frac{ \partial \mathcal{L}_X(F_{u})} {\partial F_u} & = \sum_{k} (X_{uk} - Q_{uk}) W_{kc}.
\end{split}
\]
We then update each $F_{uc}$ by gradient ascent and then project onto a space of non-negative real numbers $[0, \infty)$:
\begin{equation}
F_{uc}^{new} = \max(0, F_{uc}^{old} + \alpha ( \frac{ \partial \mathcal{L}_G(F_{u})} {\partial F_u} + \frac{ \partial \mathcal{L}_X(F_{u})} {\partial F_u}))
\label{eq:update_fuc}
\end{equation}
where $\alpha$ is a learning rate which we set using backtracking line search~\cite{Boyd04Convex}.

\xhdr{Updating logistic parameters}
We update parameters $W$ of the logistic model by keeping community memberships $F$ fixed. To compute this, we first notice that we can ignore $\mathcal{L}_G$ in Eq.~\ref{eq:mle_full}, as $G$ does not depend on $W$. Next, we also include $l_1$-regularization on $W$, as we aim to learn \emph{sparse} relationships between community memberships and node attributes:
\[
\hat{W} = \argmax_{W} \sum_{u, k}  \log P(X_{uk}|F, W) - \lambda |W|_1.
\]
Furthermore, as we employ an independent logistic model for each attribute, we only need to consider the $k$-th attribute when updating the weight vector $W_k$:
\begin{equation}
\argmax_{W_k} \sum_{u} \log P(X_{uk}|F, W_k) - \lambda |W_k|_1.
\label{eq:mle_wk}
\end{equation}
Note that this is $l_1$-regularized logistic regression with input features $F$ and output variable $X$. Again, we simply apply a gradient ascent method:
\[
\frac{\partial \log P(X_{uk}|F, W_k)} {\partial W_{kc}}  = (X_{uk} - Q_{uk}) F_{uc},
\]
\[
W_{kc}^{new} = W_{kc}^{old} + \alpha ( \sum_u \frac { \partial \log P(X_{uk}|F, W_k) } {\partial W_{kc}} - \lambda \cdot \mbox{Sign}(W_{kc})),
\]
where $\alpha$ is a step size as in Eq.~\ref{eq:update_fuc}.

Now, we iteratively update $F_u$ for each $u$ and then update $W_k$ for each attribute $k$. We stop iterating once the likelihood does not increase (by at least $0.001\%$) after a full iteration over all $F_u$ and all $W_k$.

\xhdr{Determining community memberships}
After learning real-valued community affiliations $\hat{F}$, we need to determine whether node $u$ belongs to community $c$. To do so, we regard $u$ as belonging to $c$ only if the corresponding $F_{uc}$ is above the threshold $\delta$. 
We set $\delta$ so that a node belongs to community $c$ 
if the node is connected to other members of $c$ with an edge probability higher than $1/N$.
To determine $\delta$, we need to solve:
\[
\frac{1} {N} \leq 1 - \exp(- \delta^2).
\]
Solving this inequality, we set the value of $\delta = \sqrt{- \log (1 - 1/ N)}$. We have also experimented with other values of $\delta$ and found that this value of $\delta$ gives good performance in practice.

\xhdr{Choosing the number of communities}
To automatically find the number of communities $C$, we adopt the approach used in \cite{airoldi07blockmodel}. We reserve 10$\%$ of node pairs in the adjacency matrix and node-attribute pairs as a holdout set. Varying $C$, we fit the \model with $C$ communities on 90$\%$ of node-node pairs and node-attribute pairs and then evaluate the likelihood of \model on the holdout set. The $K$ that induces the maximum held-out likelihood will be chosen as the number of communities.

\hide{When the network is too small (\eg, has less than 50 edges), we choose $C$ so as to achieve the smallest value of the Bayesian Information Criterion as in \cite{airoldi07blockmodel}:
\[
\mathit{BIC}(C) = - 2 l(\hat{F}, \hat{W}) + (NC + K(C + 1))  \log (NK + N^2),
\]
where $l$ is the log-likelihood, $NC + K(C + 1)$ is the number of model parameters (when there are $C$ communities) and $NK + N^2$ is the number of observations ($NK$ node attributes and $N^2$ elements of the adjacency matrix).
} 

\xhdr{Computational complexity of \model}
We next analyze the computational complexity of \model. In particular, we show that a full iteration of \model takes time \emph{linear} in the number of edges and attributes.

For simplicity, let us assume a single community $C = 1$, then updating $F_u$ for a single $u$ takes $N + K$ operations when computed in a naive way.
However, we can compute $\frac{ \partial \mathcal{L}_G(F_{u})} {\partial F_u}$  in $O(|\mathcal{N}(u)|)$. This means that the number of operations required to compute the gradient is proportional to the degree of node $u$ since~\cite{morup11infinite,jaewon13agmfast}:
\[
\sum_{v \not\in \mathcal{N}(u)} F_{vc} = (\sum_v F_{vc} - F_{uc} - \sum_{v \in \mathcal{N}(u)} F_{vc}).
\]
By storing $\sum_v F_{vc}$, the second term in $\frac{ \partial \mathcal{L}_G(F_{u})} {\partial F_u}$ can be computed in $O(|\mathcal{N}(u)|)$.
Therefore, updating $F_u$ for all nodes $u$ takes $O(|E| + NK)$ operations.
Because updating $W_k$ takes just $O(N)$ for each $k$, a full iteration of \model takes $O(|E| + NK)$ operations, which is linear in the number of edges, nodes and the number of attributes.

Notice that \model nicely lends itself to parallelization. In particular, updating $W_k$ naturally allows for parallelization, as we can update $W_k$ for multiple attributes $k$ simultaneously. Because $F$ is fixed, the problems in Eq.~\ref{eq:mle_wk} are independent for different attributes $k$.
We also update $F_u$ for multiple nodes $u$ in parallel. In this case, updating each $u$ is not necessarily independent for different nodes $u$. However, as shown by Niu et al.~\cite{Niu11HogWild}, updating $F_u$ in parallel works well in practice, as networks tend to be \emph{sparse}.
As we show in the next section, parallelization on a single shared memory machine boosts the speed of \model by a factor of 20 (the number of threads).

A parallel C++ implementation of \model algorithm is available as a part of the Stanford Network Analysis Platform (SNAP): \url{http://snap.stanford.edu/snap}.

\xhdr{\model hyperparameter settings}
To initialize $F$, we use locally minimal neighborhoods~\cite{Gleich12neighborhoods}. A neighborhood $N(u)$ of a node $u$ 
is {\em locally minimal} if $N(u)$ has lower conductance than all neighborhoods $N(v)$ of $u$'s neighbors $v$. Locally minimal neighborhoods have been shown to be a good initialization for community detection methods~\cite{Gleich12neighborhoods}. 

Last, notice that the overall model likelihood is a combination of the network likelihood $\mathcal{L}_G$ and the likelihood of node attributes $\mathcal{L}_X$ (Eq.~\ref{eq:mle_full}). As the two likelihoods can have vastly different ranges we scale them using the parameter $\alpha$. In particular, we introduce a hyperparameter $\alpha$ that controls the scaling between the two likelihoods:
\[
\argmax_{F \geq 0, W} (1 - \alpha) \mathcal{L}_G + \alpha \mathcal{L}_X - \lambda |W|_1.
\]
We choose values of hyperparameters $\alpha$ and $\lambda$ among $\alpha \in \{0.25, 0.5, 0.75\}, \lambda \in \{0.1, 1.0\}$ based on the held-out data likelihood (\ie, by cross-validation). We note that the performance of \model does not change much with the values of hyperparameters. Setting $\alpha = 0.5$ (\ie, the unscaled version of Eq.~\ref{eq:mle_full}) and $\lambda = 1$ gives reliable performances in most cases.

\section{Experimental Evaluation}
\label{sec:experiments}

We quantify the performance of \model by comparing it to state-of-the-art community detection methods in various social and information networks. We evaluate the performance of the methods by evaluating the accuracy of the detected communities when compared to the gold-standard, ground-truth communities. We also evaluate the scalability by measuring the running time as the network size grows. \shortver{(See the extended version~\cite{jaewon13attributes}.)}

\xhdr{Dataset description}
For our evaluation, we consider five datasets where we have network information as well as node attributes. In addition to networks and attributes, we also have access to explicit \emph{ground-truth} community labels. The availability of such ground-truth allows us to evaluate community detection methods by quantifying the degree of agreement between the detect and the ground-truth communities~\cite{Ruan13CODICIL}. Table~\ref{tab:data} lists the networks and their properties.

The networks come from 3 different domains: information network among Wikipedia articles (philosophers)~\cite{Ahn10LinkCommunitiesNature}, content-sharing network (Flickr)~\cite{Ruan13CODICIL}, and ego-networks from online social network services (Facebook, Google+, and Twitter)~\cite{julian12circles}. We next describe each of these networks in further detail.

The philosophers network~\cite{Ahn10LinkCommunitiesNature} consists of Wikipedia articles about famous philosophers. Nodes represent Wikipedia articles about philosophers, and undirected edges indicate whether one article links to another. For the attributes of each node $u$, we use a binary indicator vector of out-links from node $u$ to other non-philosopher Wikipedia articles. For example, we regard a link to a Wikipedia article ``Edinburgh'' as a binary attribute ``Edinburgh.'' We consider 5,770 attributes, to which at least five philosophers have a link.  Moreover, Wikipedia also provides categories (\eg, ``Muslim philosophers'', or ``Early modern philosophers'') for each article. We regard each category with more than five philosophers as a ground-truth community.

The Flickr image sharing network~\cite{Ruan13CODICIL} consists of nodes which represent Flickr users, and edges indicate follow relations between users. We use tags of images uploaded by a given user as her attributes. In this network, the ground-truth communities are defined as user-created interest-based groups that have more than five members.

The last three networks (Facebook, Google+, and Twitter) are ego-networks that are available from the Stanford Large Network Dataset Collection (\url{http://snap.stanford.edu/data}). To obtain ground-truth communities and node attributes, we use the same protocol as in \cite{julian12circles}. Ground-truth communities are defined by social circles (or ``lists'' in Twitter), which are manually labeled by the owner of the ego-network. In Facebook and Google+, node attributes come from user profiles, such as gender, job titles, institutions, and so on. In Twitter, node attributes are defined by hashtags used by the user in her tweets. To reduce the dimensionality of the node attributes, we discard any attribute which the owner of the ego-network does not possess.

\begin{table}[t]
\centering
    \begin{tabular}{lrrrrrr}
    \hline
    Dataset & $N$ & $E$ & $C$ & $K$ & $S$ & $A$ \\ \hline 
    Facebook &4,089 & 170,174 & 193 & 175 & 28,76 & 1.36\\ 
    Google+ & 250,469 & 30,230,905 & 437 & 690 & 143.51 & 0.25\\ 
    Twitter & 125,120 & 2,248,406 & 3,140 & 33,569 & 15.54 & 0.39\\ 
    Philosophers & 1,218& 5,972 & 1,220 & 5,770 & 6.86 & 6.87\\
    Flickr& 16,710& 716,063 & 100,624 & 1,156 & 28.91 & 174.08\\ \hline
    \end{tabular}
    \caption{Dataset statistics. $N$: number of nodes, $E$: number of edges, $C$: number of communities, $K$: Number of node attributes, $S$: average community size, $A$: community memberships per node.
    }
    \label{tab:data}
  \vspace{-7mm}
\end{table}

\xhdr{Baselines for comparison}
We consider the three classes of baseline community detection methods: (1) methods that use only the network structure, (2) methods that user only node attributes, and  (3) methods that combine both.

The first class of baselines considers only the network, ignoring node attributes altogether:
\emph{Demon}~\cite{Coscia13DEMON} and \emph{BigCLAM}~\cite{jaewon13agmfast} are state-of-the-art overlapping community detection methods.

Second is a class of baselines that focuses on node attributes without considering the network structure. Here, we use \emph{Multi Assignment Clustering (MAC)}~\cite{Frank12MultiAssignmentClustering}, which detects overlapping communities based on node attributes alone.

The third class of baselines we consider combines the network structure with node attributes. For this class, we choose three state-of-the-art methods. Based on Table~\ref{tab:baselines} we select one algorithm from each model type:
\emph{Block-LDA}~\cite{Cohen11BlockLDA} represents soft-membership approaches,
%
while the \emph{CODICIL}~\cite{Ruan13CODICIL} represents heuristics for non-overlapping communities, and the \emph{EDCAR}~\cite{Gunnermann13EDCAR} represents heuristics for finding dense subgraphs.
Finally, we consider the \emph{\julian}~\cite{julian12circles} method, which represents overlapping hard-membership approaches.

\hide{
The third class of baselines combines the network structure with node attributes:
\emph{CODICIL} \cite{Ruan13CODICIL} is a technique that combines network edges and node attributes and can be combined with a community detection method; we considered the implementations from \cite{Ruan13CODICIL}, which use Metis~\cite{karypis98metis} and Markov clustering~\cite{MarkovClustering2009}; we show results with Markov clustering (which outperformed Metis).
Also from this class we consider \emph{Block-LDA} \cite{Cohen11BlockLDA}, a variant of Latent Dirichlet Allocation (LDA)~\cite{blei03lda} which combines the network information and node attributes (`words'); we regard each node attribute as a word occurrence.
Since Block-LDA estimates ``soft'' (\ie, real-valued) community memberships, we convert them into hard memberships by thresholding.
Finally from this class we consider McAuley and Leskovec (\emph{\julian})~\cite{julian12circles}, which is a state-of-the-art method for overlapping community detection with node attributes.
We use publicly available implementations of each of the above methods.

Note that the above baselines represent the current state-of-the-art in a wide range of community detection scenarios: Our baselines include methods that detect non-overlapping communities (CODICIL), methods that estimate stochastic memberships (\ie, probabilities associated with each community) such as Block-LDA, and methods that detect overlapping communities (BigCLAM, Demon, \julian).
In particular, Block-LDA represents a class of LDA-like models~\cite{blei03lda, LinkLDA}, which consider words in individual documents (node attributes in our framework) in addition to relationships between the documents (the network).
Finally, we note that we also considered other baselines~\cite{rosvall08infomap, Ahn10LinkCommunitiesNature,airoldi07blockmodel}\cite{Gunnermann13EDCAR}, including those that make use of node features only \cite{hierarchical}; we mention briefly that none of these alternatives outperformed \model on average.
}

For all baselines, we use implementations provided by the authors. All baselines except CODICIL require a user to specify the number of communities to detect. 
We set this parameter so that each model detects the same number of communities as \model. 
CODICIL and EDCAR also has other input parameters, for which we used default values provided by the authors.

\xhdr{Evaluation metrics}
We quantify the performance in terms of the agreement between the ground-truth communities and the detected communities. To compare a set of ground-truth communities $C^*$ to a set of detected communities $C$, we adopt an evaluation procedure previously used in \cite{jaewon13agmfast}: Every detected community is matched with its most similar ground-truth community. Given this matching, we then compute the performance. We also then take every ground-truth community and match it with a detected community and again compute the performance. Our final performance is the average of these two metrics. We average the two scores because matching only from one side leads to degenerate optimal performance (for example, outputting all possible subsets of nodes as detected communities would achieve perfect matching ground-truth communities to the detected ones).

More formally, our evaluation function is:
\begin{equation}
 \frac{1}{2|C^*|} \sum_{C^*_i \in C^*} \max_{C_j \in C}\delta(C^*_i, C_j) + \frac{1}{2|C|} \sum_{C_j \in C} \max_{C^*_i \in C^*}\delta(C^*_i, C_j),
 \label{eq:metric}
\end{equation}
where $\delta(C^*_i, C_j)$ is some similarity measure between the communities $C^*_i$ and $C_j$. We consider two standard metrics $\delta(\cdot)$ for quantifying the similarity between a pair of sets, namely the $F1$ score and the Jaccard similarity. Thus, for each method, we obtain a score between 0 and 1, where 1 indicates the perfect recovery of ground-truth communities.

\begin{table*}[t]
\small
\centering
    \begin{tabular}{l|c|ccccc|ccccc|c}
    \hline
                                 & &\multicolumn{5}{c|}{$F1$ score}               & \multicolumn{5}{c|}{Jaccard similarity}  \\
    Method&Info&Phil&Flickr& Facebook&Google+&Twitter &Phil&Flickr& Facebook&Google+&Twitter& Avg.\\ \hline
    Demon & Net &0.244$^*$&0.171$^*$&0.386$^*$&0.323$^*$&0.280$^*$&0.143$^*$&0.098$^*$&0.283$^*$&0.234&0.186$^*$&0.235$^*$\\
    BigCLAM & Net &0.276$^*$&0.166$^*$&0.455&0.341&0.359$^*$&0.156$^*$&0.092$^*$&\textbf{0.347}\hphantom{$^*$}&0.231&0.246$^*$&0.267$^*$\\
    MAC & Attr &0.117$^*$& N/A&0.297$^*$&0.159$^*$&0.246$^*$&0.069$^*$& N/A&0.190$^*$&0.101$^*$&0.154$^*$&0.133$^*$\\
    Block-LDA & Both &0.146$^*$& N/A&0.356$^*$&0.307&0.273$^*$&0.082$^*$& N/A&0.241$^*$&0.204$^*$&0.173$^*$&0.178$^*$\\
    CODICIL & Both &0.277$^*$&0.132$^*$&0.378$^*$&0.247$^*$&0.279$^*$&0.167$^*$&0.079$^*$&0.263$^*$&0.166$^*$&0.190$^*$&0.218$^*$\\
    EDCAR & Both & 0.264$^*$ & 0.112$^*$ & 0.321$^*$ & 0.135$^*$ & 0.258$^*$ & 0.157$^*$ & 0.051$^*$ & 0.222$^*$ &  0.081$^*$ & 0.165$^*$ & 0.177$^*$ \\
    \julian & Both & N/A& N/A&0.401$^*$&\textbf{0.365}\hphantom{$^*$}&0.319$^*$& N/A& N/A&0.265$^*$&\textbf{0.254}\hphantom{$^*$}&0.211$^*$&0.183$^*$\\
    \model & Both &\textbf{0.314}\hphantom{$^*$}&\textbf{0.183}\hphantom{$^*$}&\textbf{0.462}\hphantom{$^*$}&0.352\hphantom{$^*$}&\textbf{0.362}\hphantom{$^*$}&\textbf{0.192}\hphantom{$^*$}&\textbf{0.106}\hphantom{$^*$}&\textbf{0.347}\hphantom{$^*$}&0.249\hphantom{$^*$}&\textbf{0.249}\hphantom{$^*$}&\textbf{0.282}\hphantom{$^*$}\\ \hline
    \end{tabular}
    \caption{Performance of methods on five datasets. {\em Info} indicates the information used by a given method (Network, Attributes, or Both). Best performing models are bolded.
    Symbol $^*$ indicates that \model outperforms a given baseline by $95\%$ statistical confidence.
    Overall, \model statistically significantly outperforms all considered methods.
    }
    \label{tab:acc}
    \vspace{-7mm}
\end{table*}

\xhdr{Experiments on recovering ground-truth communities}
We evaluate the performance of \model and baselines on our five datasets.  Table~\ref{tab:acc} shows the results where ``N/A'' means that the method cannot scale to a given network. We make several observations.

Comparing \model to methods without the node attributes (Demon and BigCLAM), we notice that \model achieves better performance, as it combines the information from the node attributes as well as the network. Similarly, \model also outperforms MAC, which only focuses on node attributes. In particular, \model never performs worse than state-of-the-art methods that use only a single source of data. The strong performance of \model is not obvious, as it would be entirely possible that combining two sources of data would confuse the algorithm and degrade the overall performance (in fact, notice that BigCLAM, which uses only the network structure, indeed outperforms most of the methods that use both sources of information). Thus, we believe that the strong performance of \model as an indication that \model combines the best ingredients from both worlds.

When comparing the performance of \model to methods that consider both the network structure and node attributes (CODICIL, Block-LDA, and \julian), we again observe the strong performance of \model. On average, \model gives 47\% relative improvement in the accuracy of detected communities over methods that consider both sources of information.

%

We also note that \model shows a bigger margin in performance against the baselines in an information network such as the philosophers dataset, or a content-sharing network like Flickr than in social networks. In the philosophers network, for example, \model achieves a 14\% relative gain in the $F1$ score and 15\% in the Jaccard similarity compared to the best baseline. A possible explanation for this phenomenon is that in content-sharing and information networks, the properties/content of the nodes plays a much bigger role in link formation.

Overall, we note that across all datasets and evaluation metrics, \model yields the best performance in 8 out of 10 cases. In terms of average performance, \model outperforms Demon by 20\%, BigCLAM by 6\%, MAC by 112\%, Block-LDA by 58\%, CODICIL by 29\%, EDCAR by 57\%, and \julian by 54\%.

Last, we also measure the statistical significance of performance differences of \model and the baselines. For each baseline's performance on each data set, we compute the statistical significance of \model outperforming the baseline using a one-sided $Z$-test. We use the symbol $^*$ in Table~\ref{tab:acc} to indicate a 95\% statistical significance level.  On the philosophers, Flickr, and Twitter datasets, \model outperforms every baseline at a $95\%$ significance level. On Facebook, \model outperforms all baselines, at a $95\%$ significance level in all but one case. On Google+, \model performs the second best compared to \julian.

%

\xhdr{Experiments on partially observed networks}
Combining network and attribute information into a single method should, in principle, lead to the development of a more robust community detection algorithm. In particular, when networks may be incomplete or partially observed, the performance of \model should degrade gently, as it should be able to rely on the node attribute information; this way, it should compensate for the noise in the network structure.


To investigate the robustness of performance under an unreliable network structure, we next explore the problem of detecting communities from partially observed networks where some fraction of edges are missing while the node attributes are fully available. 
For the sake of evaluation, we remove a fraction $\gamma$ of edges in the network uniformly at random.
Note that we regard a removed edge in the same way as an unobserved edge, because in practice we cannot distinguish between edges that do not exist (\eg, users who aren't friends) and edges that are unobserved (\eg, users who haven't gotten around to declaring their friendship yet).

Rather than examining performance of all 6 baselines, we focus on making a comparison over the three top baselines that use either the network or the node attributes: BigCLAM, which considers the network only and is the best baseline in our experiments; MAC, which only considers the node attributes; and CODICIL, which is the best performing baseline that considers both the network and the attributes. For each baseline, we measure the relative performance that \model achieves over the baseline:
\[
\frac{ F1^\gamma\mbox{(\model)} - F1^\gamma (\mbox{Baseline})} {F1^\gamma (\mbox{Baseline})}
\]
where $F1^\gamma$ is the $F1$ score in Eq.~\ref{eq:metric} for the network with $\gamma$ fraction of edges removed.

\begin{figure}[t]
\centering
  \subfigure[Google+]{\label{fig:edge.del.gp}\includegraphics[width=0.23\textwidth]{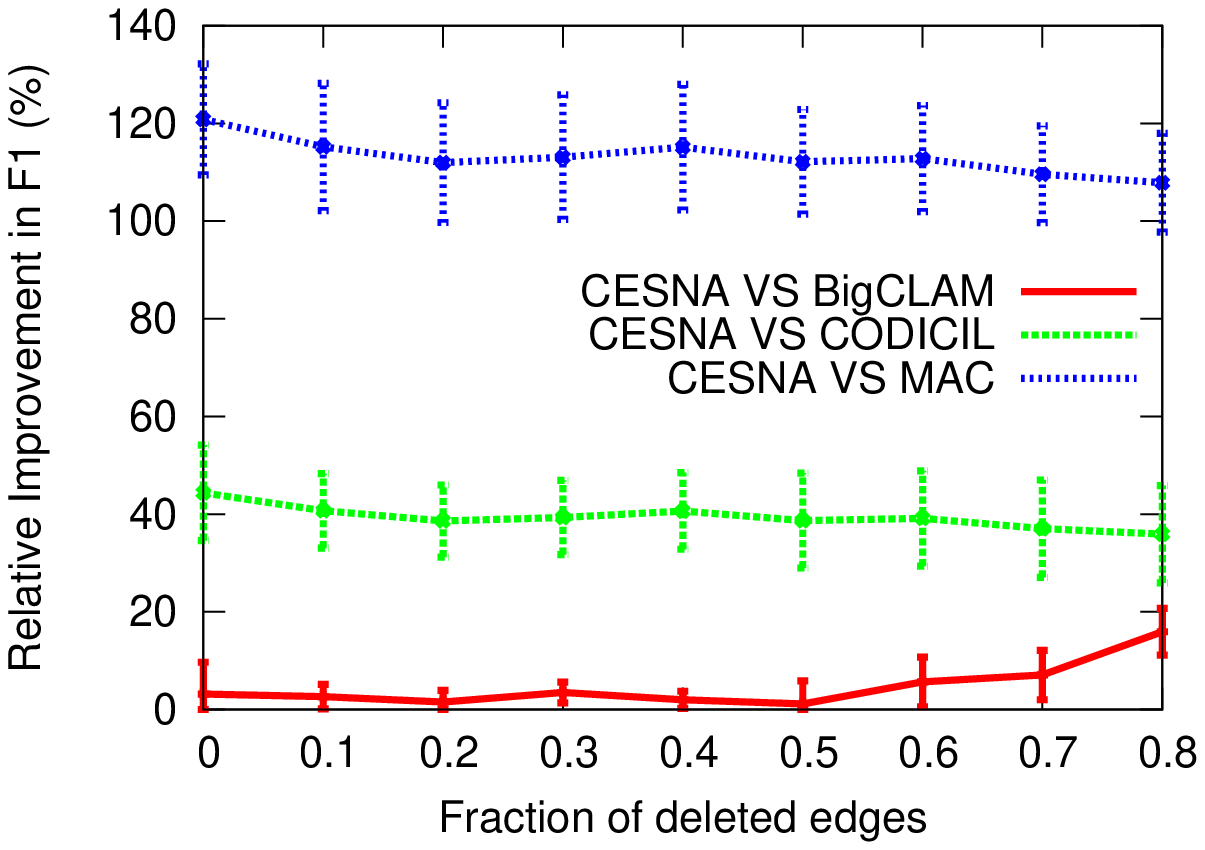}}
  \subfigure[Facebook]{\label{fig:edge.del.fb}\includegraphics[width=0.23\textwidth]{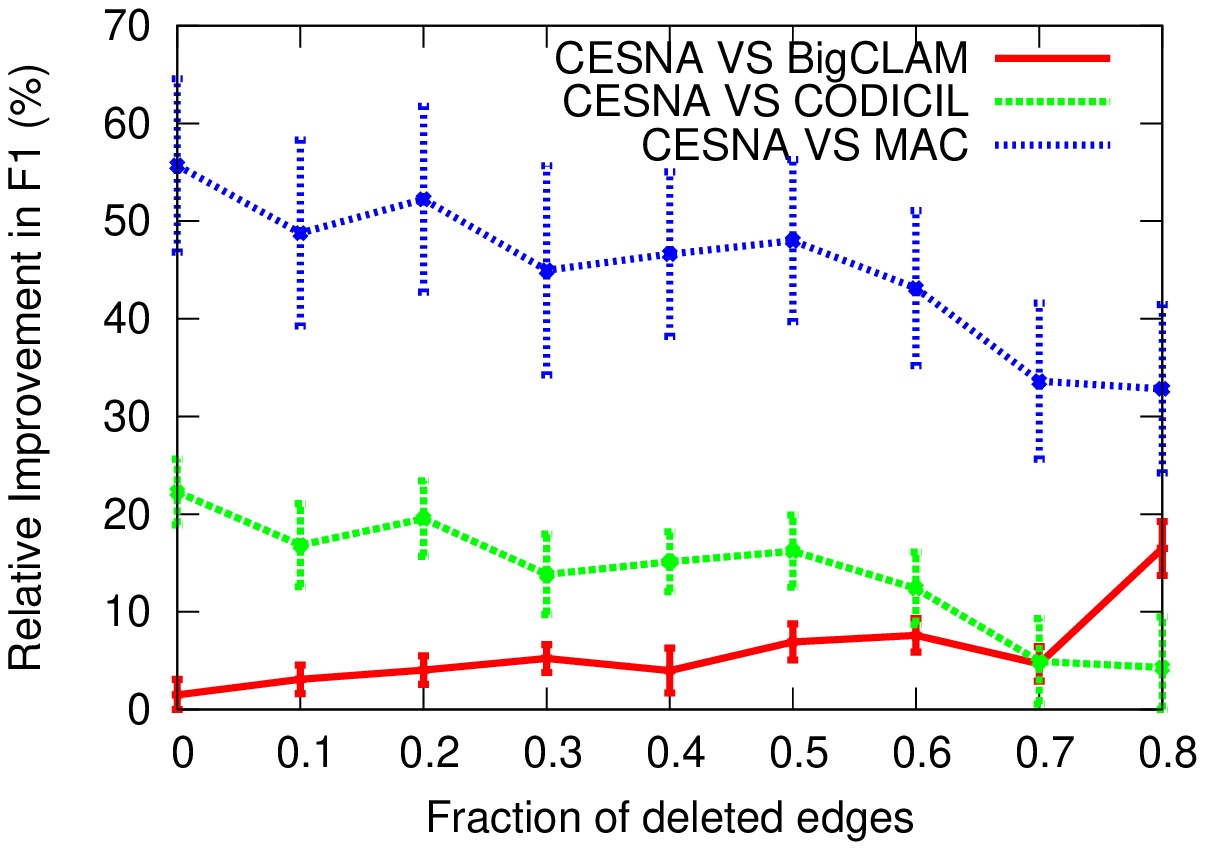}}
  \subfigure[Twitter]{\label{fig:edge.del.tw}\includegraphics[width=0.23\textwidth]{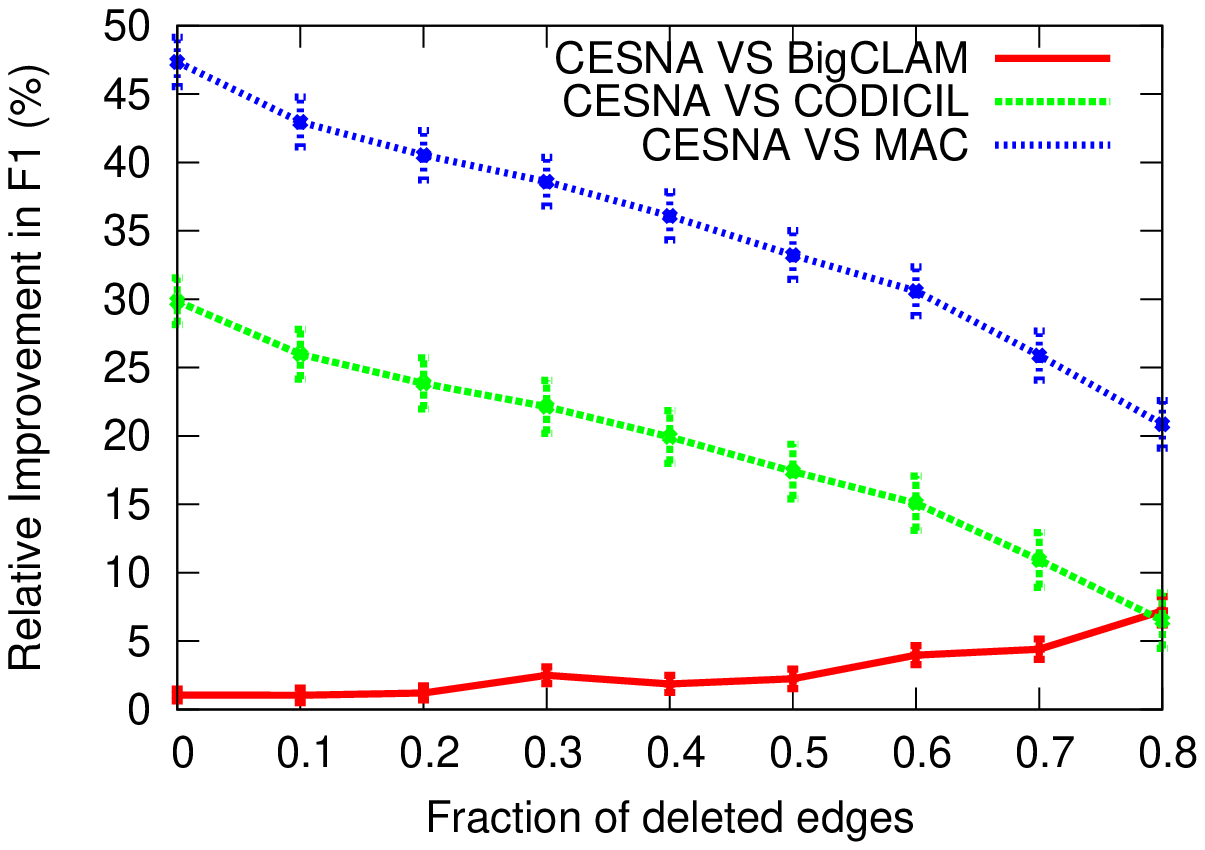}}
  \subfigure[Flickr]{\label{fig:edge.del.phil}\includegraphics[width=0.23\textwidth]{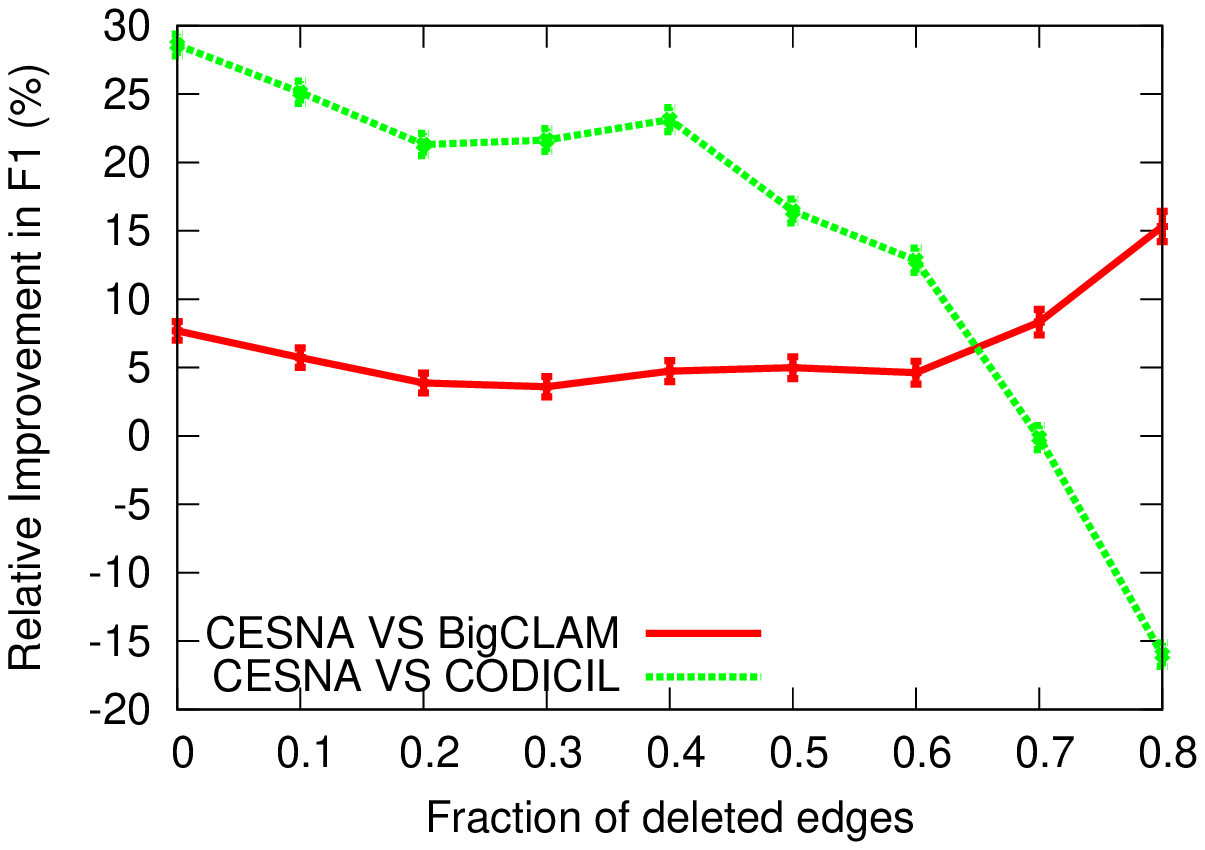}}
   \caption{Relative gain in F1 over the best method with network information only (BigCLAM), the method with node attributes only (MAC), and the method with both network and attributes (CODICIL)  
   when edges are randomly removed.}
   \label{fig:edge.del}
   \vspace{-3mm}
\end{figure}

In Figure~\ref{fig:edge.del}, we display experimental results (with standard deviation) as we vary from $\gamma = 0$ to $\gamma = 0.8$. We consider all datasets except philosophers (for which, results are too noisy due to the small network size).
For Flickr, we omit performance of MAC, as the algorithm was not able to process it due to too high time and space complexity.

In all cases, we note similar behavior (Figure~\ref{fig:edge.del}). As the network becomes more unreliable, the improvement of \model over BigClam increases. On the other hand, for methods that use node attributes (and the network structure), we note that in Google+, the performance improvement of \model remains constant, while in Facebook and Twitter, the performance improvement of \model slowly shrinks as more and more of the network structure gets removed.

The results are intuitive: Even though the network contains many missing edges, \model still outperforms other methods by better leveraging the information present in the node attributes.
The results with MAC and CODICIL, which are decreasing functions of $\gamma$, nicely shows that the performance gain from the network structure diminishes as we remove more edges.

Last, we also briefly note that similar results are observed with the relative improvement in Jaccard similarity, and that \model consistently outperforms the other four baselines not shown in Figure~\ref{fig:edge.del} for every value of $\gamma$.



\longver{
\xhdr{Evaluating scalability}
We evaluate the scalability of community detection methods by measuring each method's running time on synthetic networks as we increase the network size. Using the Forest Fire model~\cite{jure05dpl}, we generate synthetic networks with the forward and backward probabilities set to $0.36$ and $0.32$, respectively. For attributes, we generate $K = 10$ attributes for each node with independent Bernoulli random variables with probability 0.5.

Figure~\ref{fig:scalability} shows the running time of methods versus the network size. 
Among the four baselines that consider both network and the node attributes (\ie, Block-LDA, CODICIL, EDCAR, \julian), we show CODICIL since it is the fastest among the four.
We also consider a parallelized version of \model (\model (24 threads)).

Overall, we notice that \model is the second-fastest method overall, next to BigCLAM. However, we note that BigCLAM is expected to be faster than \model, as it uses a similar optimization procedure as \model yet without considering node attributes. MAC is the slowest, and CODICIL is the second-slowest method. 
DEMON is faster than \model for small networks (up to 100,000 nodes), though \model is faster when the network becomes larger.

We obtain even further speedup by considering a parallel implementation of \model. Using 24 threads on a single machine, \model takes just 10 minutes to process a 300,000 node network.

Last, we also note that all the baselines shown in Fig.~\ref{fig:scalability} solve ``simpler'' problems than \model. For example, CODICIL detects non-overlapping communities, which is simpler than detecting overlapping communities. Demon and BigCLAM consider only network information, ignoring node attributes. Nevertheless, \model is faster than CODICIL and Demon, and it takes about 30\% more time than BigCLAM. Comparing \model to methods that achieve the same goal --- that is, overlapping community detection with node attributes (\ie, Block-LDA, EDCAR, and \julian) --- \model has a considerable advantage in scalability, as it is about an order of magnitude faster.

\begin{figure}[t]
\begin{center}
\includegraphics[width=0.8\linewidth]{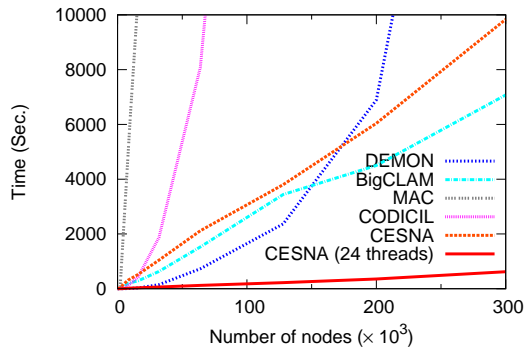}
\end{center}
\vspace{-6mm}
\caption{Algorithm runtime comparison.
Block-LDA and \julian are omitted as they took more time than 10,000 seconds for networks larger than 1\% of the X-axis (3,000 nodes).
}
\vspace{-3mm}
\label{fig:scalability}
\end{figure}
}

\section{Analysis of Detected Communities}
\label{sec:discussion}



Incorporating node attributes into community detection gives two direct advantages. The first
 advantage is the improved accuracy in community detection, which we observed in the previous section.
The second advantage is that the node attributes provide cues for interpreting detected communities.
For example, a community in a Facebook ego-network might consist of a set of high-school friends, and the homogeneity of a particular attribute in a given community might help us to interpret and explain its existence.  
Such interpretations are an important part of community detection~\cite{Ahn10LinkCommunitiesNature,airoldi07blockmodel,palla05_OveralpNature}, yet finding them is very time-consuming and may require domain knowledge, as in traditional settings, one has to infer the meaning of a given community based only on the identities of its members. By incorporating node attributes, however, \model allows us to characterize a community by examining the attributes associated with high logistic weights in the model.

In this section, we qualitatively analyze our results in the Facebook network and the philosophers network to provide insights as to how \model brings the two advantages (better interpretability and higher accuracy). In both networks, we find that \model is able to find the attributes that are naturally related to the communities. On philosophers data, we also show how \model can improve the accuracy of detected communities by incorporating node attributes.

\xhdr{Analysis of Facebook communities}
\model learns the logistic model weight $W_{kc}$ for each attribute $k$ and community $c$. Highly positive values of $W_{kc}$ mean that members of community $c$ are likely to have attribute $k$, and a highly negative value means the opposite (members are likely \emph{not} to have the attribute).
Not every attribute will be associated with community memberships, as some attributes may be irrelevant for a given community. To characterize the level of association between communities and attributes $k$, we measure the $l_2$ norm $\|W_k\|$ of its logistic weight $W_k = \{W_{k \cdot}\}$.

To examine which attributes are related to communities (either positively or negatively), we examine detected communities in Facebook ego-networks. We find that the top attributes are related to schools, including the schools attended, the types of education that users received, and the major.
On the other hand, the bottom five attributes include work start dates, work end dates, and locale. None of them act as social factors around which communities on Facebook form.


\xhdr{Analysis of Philosophers communities}
To analyze the member nodes of communities along with their related attributes, we examine the communities in the Philosophers network.

\begin{figure}[t]
\centering
  \subfigure[Islamic (\model)]{\label{fig:arab.model}\includegraphics[height=0.25\textwidth]{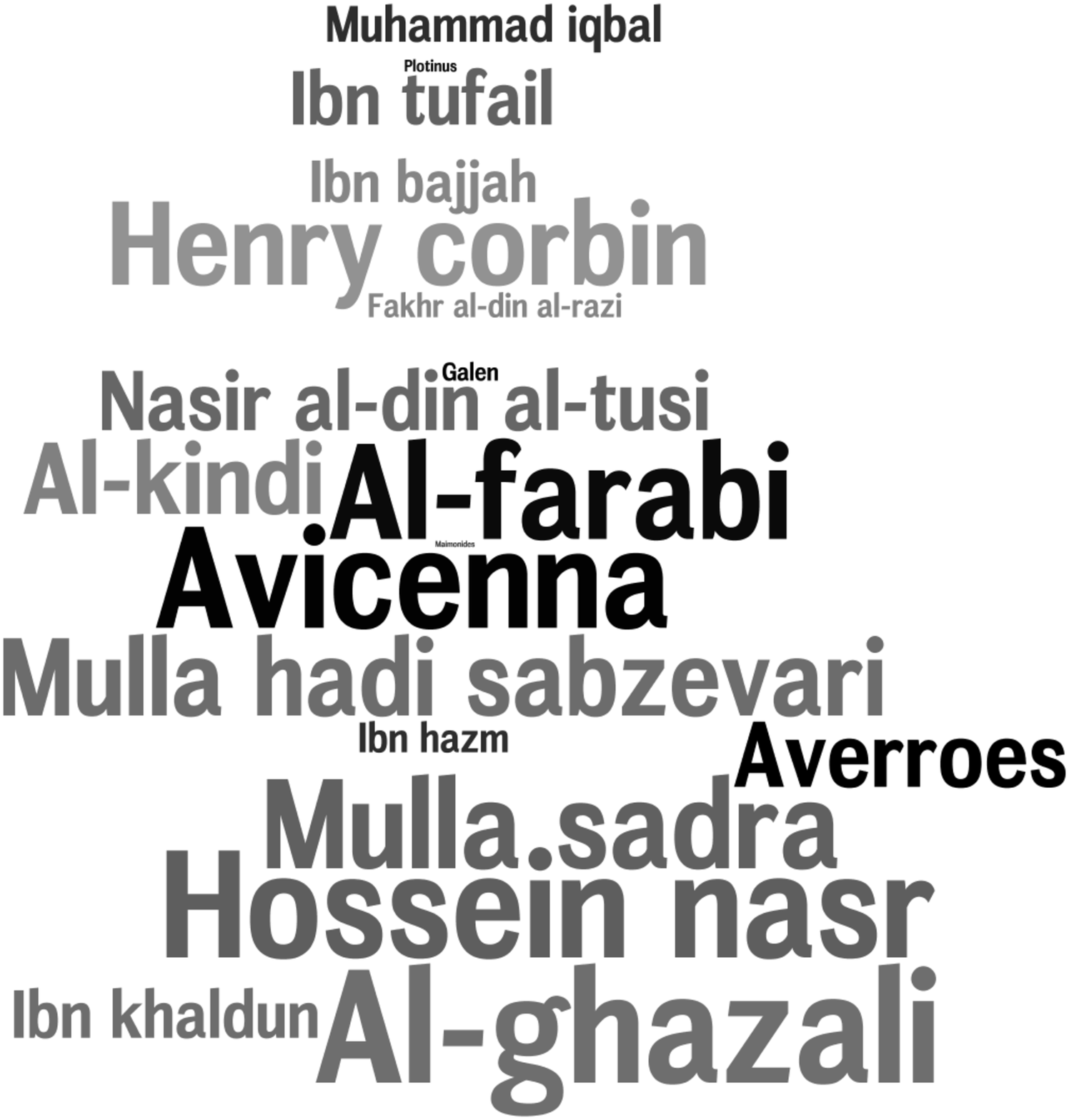}}
  \subfigure[Islamic (BigCLAM)]{\label{fig:arab.bigclam}\includegraphics[height=0.25\textwidth]{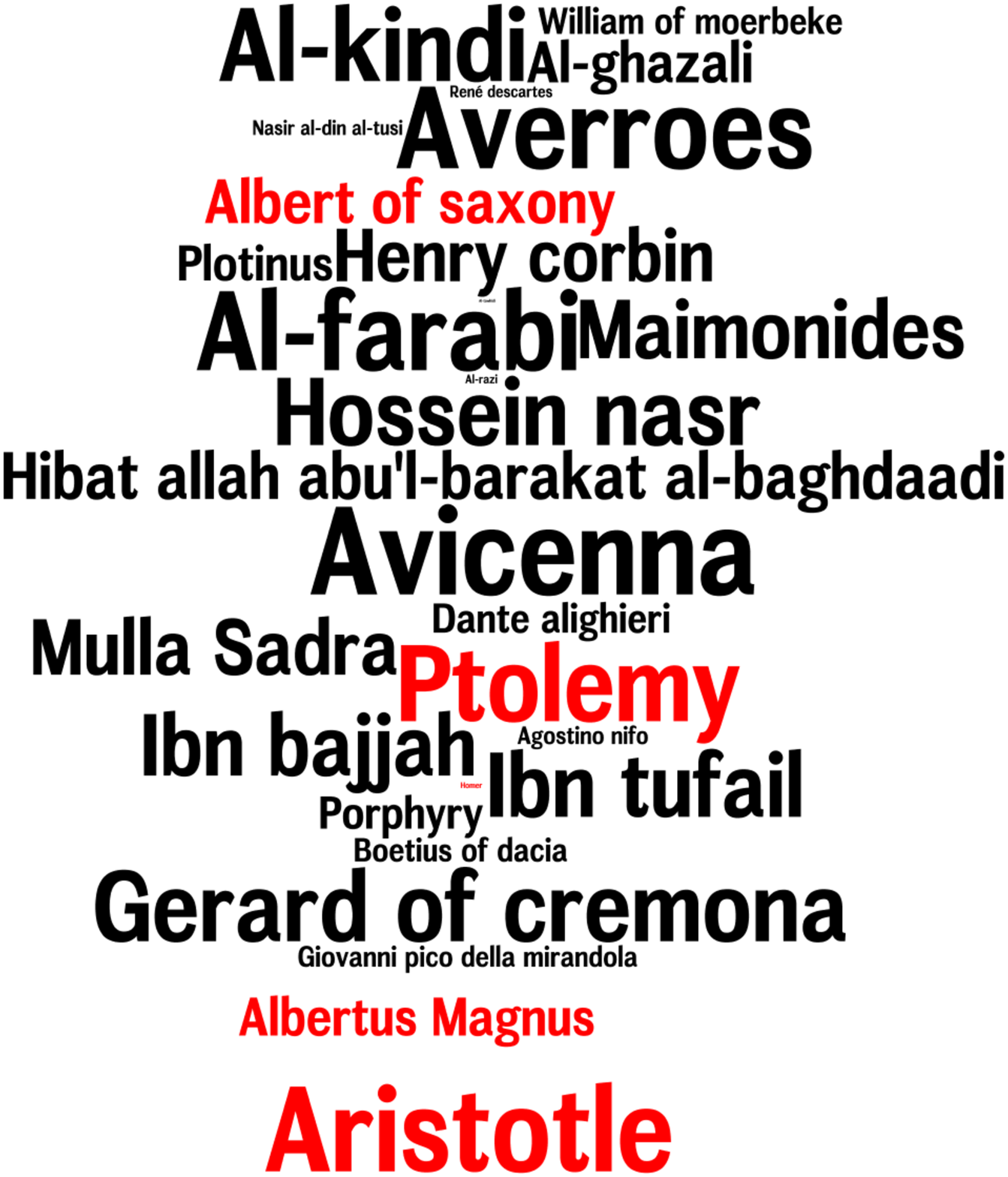}}\\
  \subfigure[Theologians (\model)]{\label{fig:theo.model}\includegraphics[height=0.25\textwidth]{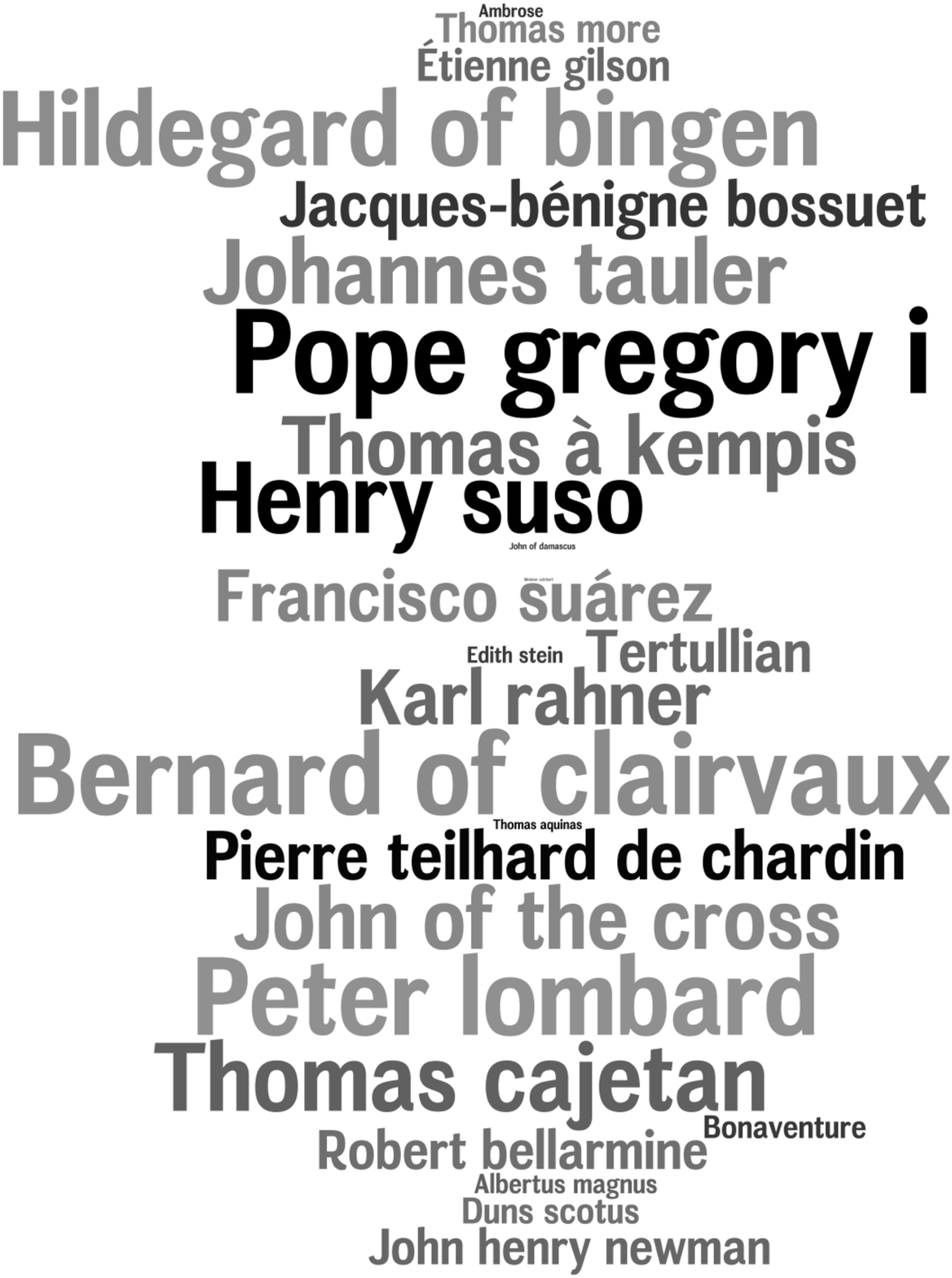}}
  \subfigure[Theologians (BigCLAM)]{\label{fig:theo.bigclam}\includegraphics[height=0.25\textwidth]{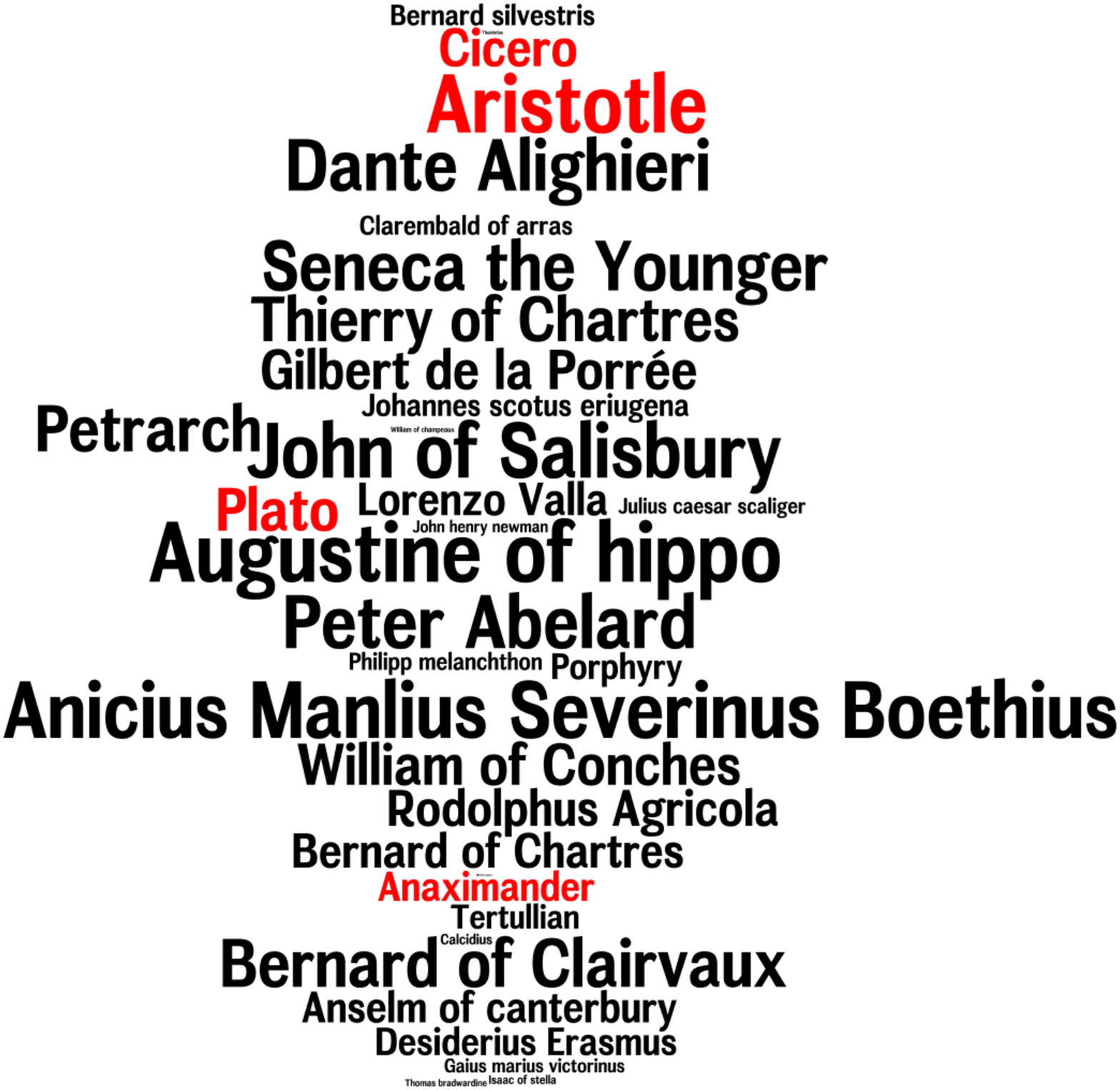}}
   \caption{
   Communities of philosophers found by \model (left) and equivalent communities detected by BigCLAM (right).
   Top: Community of Islamic philosophers, Bottom: Community of theologians,
   BigCLAM regards some notable philosophers in red letters as belonging to the communities, even though these philosophers have little to do with theology / Islam.
   \model does not make such mistakes, as \model jointly learns attributes associated with the community. (Attributes are in Fig.~\ref{fig:phil.attr}.)
   }
   \label{fig:phil.comm}
\end{figure}

\begin{figure}[t]
\centering
  \subfigure[Islamic (Attributes)]{\label{fig:arab.attr}\includegraphics[width=0.21\textwidth]{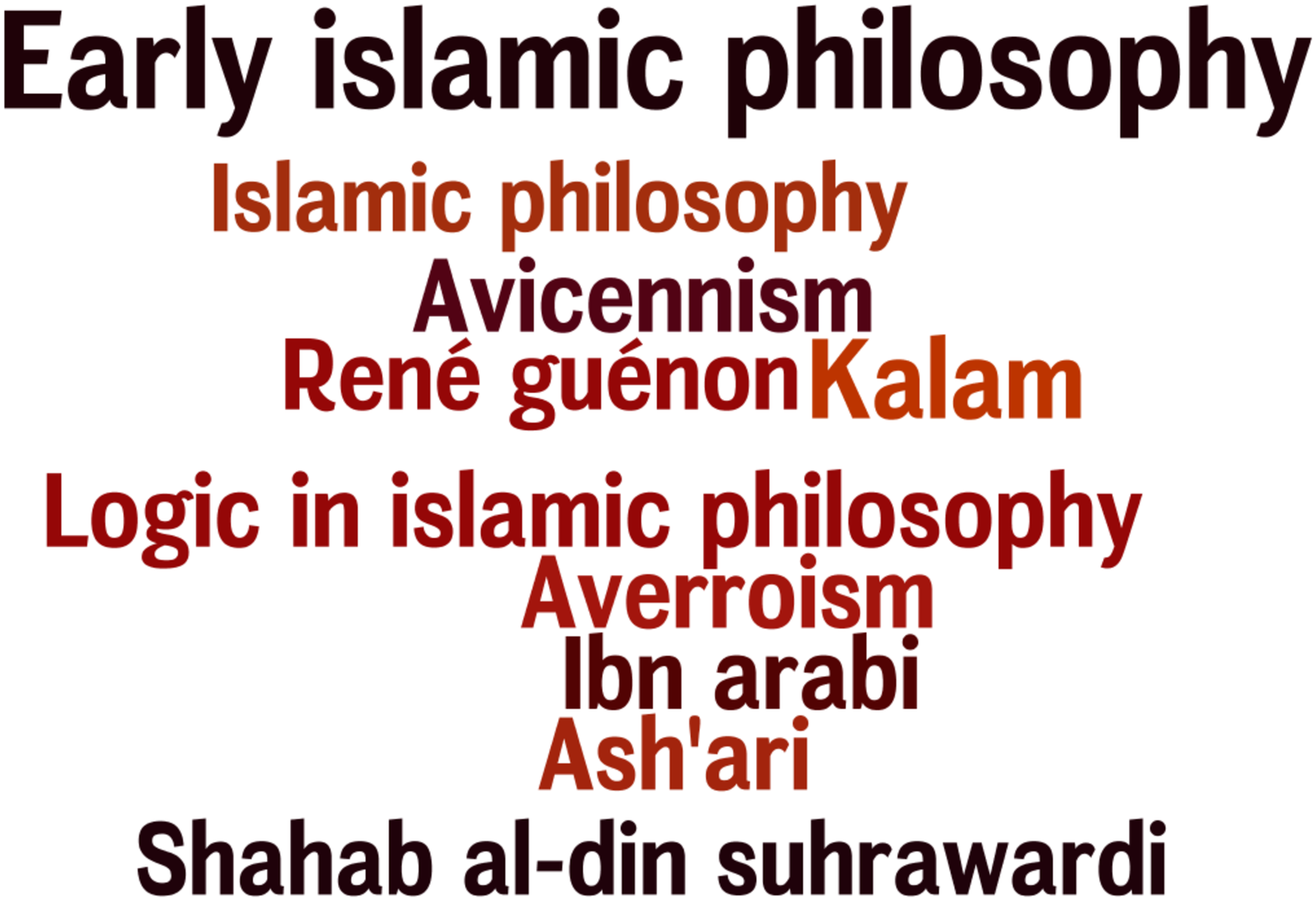}}
  \subfigure[Theologians (Attributes)]{\label{fig:theo.attr}\includegraphics[width=0.21\textwidth]{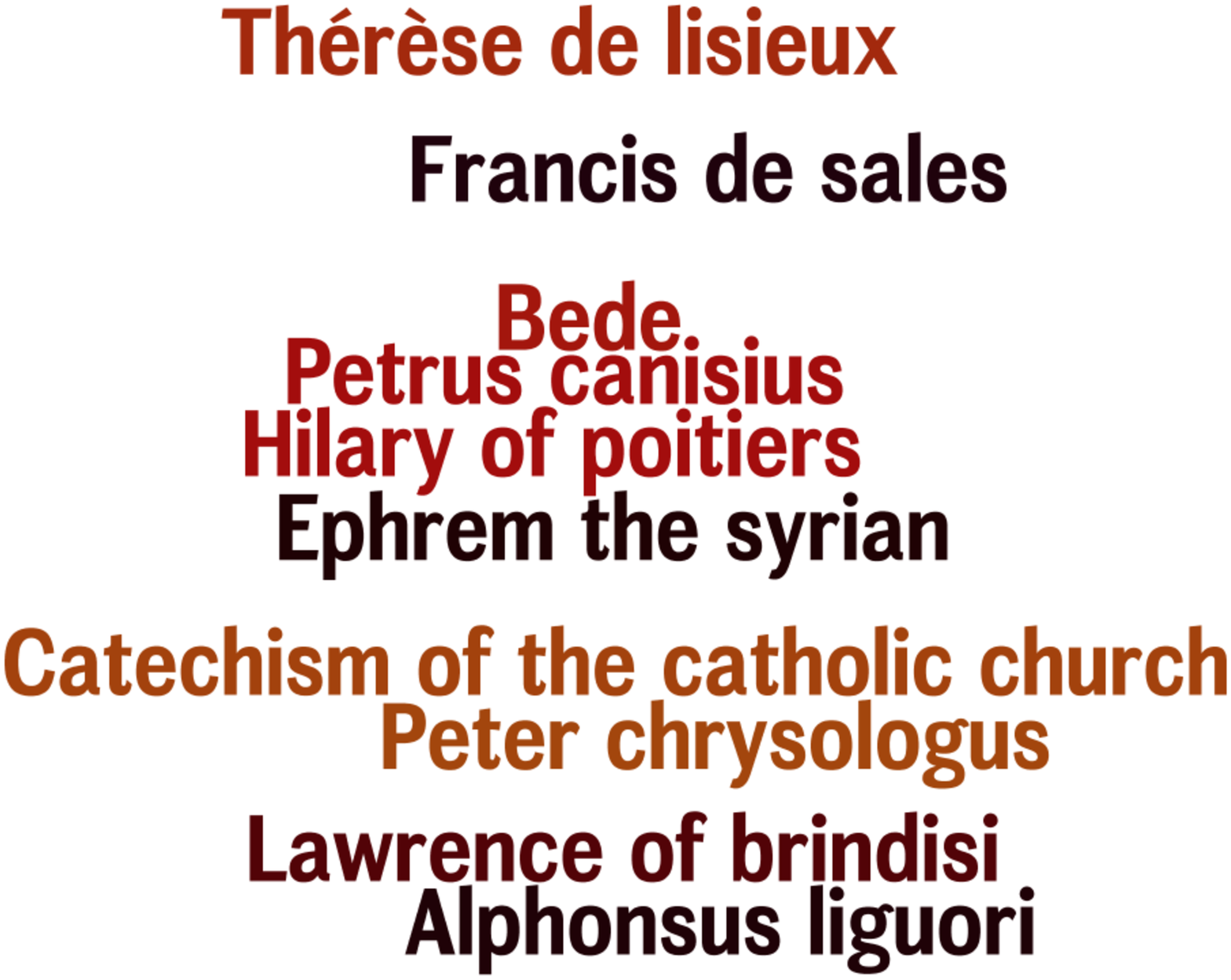}}
   \caption{
   The node attributes which \model learns to be associated with the communities.
   Left: For the community of Islamic philosophers, Right: For the community of theologians,
   }
   \label{fig:phil.attr}
\end{figure}

First, using \model, we identify communities, and then for each community we identify the top ten positively related attributes. In Figures~\ref{fig:arab.model}, \ref{fig:theo.model} we show two of the detected communities. The figure displays the titles of the corresponding Wikipedia articles. 
Moreover, we also show the attributes associated with the two communities in Figure~\ref{fig:phil.attr}.
In this figure, word sizes are proportional to the value of the logistic weight $W_{kc}$, \ie, more relevant attributes are larger.
Note that node attributes in this network represent Wikipedia articles other than philosophers to which the node links, \eg, the attributes include famous non-philosophical figures, abstract concepts, historic events, places, and so on.

First, based on the names of important attributes, \eg, ``Early Islamic Philosophy,''  we observe that the community in Figure~\ref{fig:arab.attr} represents Islamic philosophers, even without querying for the names of the philosophers in Figure~\ref{fig:arab.model}. These attributes also include some non-philosophical people related to Islam (\eg, Ren\'{e} Gu\'{e}non).

Similarly, Figures~\ref{fig:theo.model} and~\ref{fig:theo.attr} show the members of the second community detected by \model and the top ten related node attributes. Again, ``Catechism of the Catholic Church'' tells us that this community consists of theologians. The node attributes also include many priests (\eg, Lawrence of Brindisi, Bede, Hilary of Poitiers, Petrus Canisius, and Francis de Sales).

We also compare these communities to those detected by the BigCLAM. For each community detected by \model in Figs.~\ref{fig:arab.model} and~\ref{fig:theo.model}, we identify the most similar BigCLAM community based on the $F_1$ score. Figures~\ref{fig:arab.bigclam} and~\ref{fig:theo.bigclam} show these communities as detected by BigCLAM.

Interestingly, we note that the communities detected by BigCLAM contain some philosophers (in red) who are not Islamic philosophers/theologians. The reason is that these philosophers (in red) are so influential that they are very well connected to other members of the community. For example, Aristotle is connected to 229 philosophers (about one fifth of all the nodes); thus, he appears in both BigCLAM communities in Figure~\ref{fig:phil.comm}. However, by leveraging node attributes, \model does not make this mistake and finds that Aristotle does not share the same attributes as any Islamic philosophers or theologians, which, thus, excludes him.

\section{Conclusion}
\label{sec:conclusion}


In this paper, we developed \model, a scalable method for overlapping community detection in networks with node attributes. Its comparison to the state-of-the-art baselines reveals that \model exhibits improved performance both in terms of the accuracy of the detected communities as well as in scalability. \model has a linear runtime in the network size and is able to process networks an order of magnitude larger than comparable approaches. Moreover, \model also helps with the interpretation of detected communities by finding relevant node attributes for each community.

There are many possible directions for future work. One direction is to extend \model to handle more general types of attributes. Similarly, extending the method to cluster the attributes into ``topics,'' while also identifying communities would likely lead to even easier interpretation of detected communities. Finally, incorporating other sources of information than node attributes, such as information diffusion~\cite{Barbieri13CascadeCommunity} or edge attributes~\cite{Cohen11BlockLDA}, would also be possible.

\xhdr{Acknowledgements}
We thank Yiye Ruan for sharing the CODICIL code and the Flickr data.
This research has been supported in part by NSF
IIS-1016909,              
CNS-1010921,              
CAREER IIS-1149837,       
IIS-1159679,              
ARO MURI,                 
DARPA GRAPHS,             
ARL AHPCRC,
Okawa Foundation,          
PayPal, 
Docomo,                    
Boeing,                    
Allyes,                    
Volkswagen,                
Intel,                     
Alfred P. Sloan Fellowship, and    
the Microsoft Faculty Fellowship. 



\end{document}